\documentclass[12pt]{iopart}
\usepackage[utf8]{inputenc}
\expandafter\let\csname equation*\endcsname\relax
\expandafter\let\csname endequation*\endcsname\relax

\usepackage{lipsum}
\usepackage{braket}
\usepackage{graphicx}
\usepackage{dcolumn}
\usepackage{enumitem}
\usepackage{bm}
\usepackage{amsmath}
\usepackage{amssymb}
\usepackage{mathtools}
\usepackage{subfigure}
\usepackage{color}
\usepackage{xcolor}
\usepackage{verbatim}
\usepackage{float}

\newcommand{\abs}[1]{\lvert #1 \rvert} 
\newcommand{\ev}[1]{\langle #1 \rangle} 
\newcommand{\lv}[0]{\mathcal{L}} 
\newcommand{\td}[1]{\tilde{#1}} 
\newcommand{\mbf}[1]{\mathbf{#1}} 
\newcommand{\mD}[0]{\mathcal{D}} 
\newcommand{\kE}[0]{k_\text{eff}} 
\newcommand{\nif}[1]{\left\lvert\left\lvert #1 \right\rvert\right\rvert} 
\newcommand{\mJ}[0]{\mathcal{J}} 
\newcommand{\nUp}[0]{n_\uparrow} 
\newcommand{\nDown}[0]{n_\downarrow} 
\newcommand{\hSig}[0]{\hat{\sigma}} 
\newcommand{\gE}[0]{g_{\text{eff}}} 
\newcommand{\tdq}[0]{\tilde{q}} 

\begin{document}

\title[Squeezed state metrology with Bragg interferometers]{Squeezed state metrology with Bragg interferometers operating in a cavity}

\author{Athreya Shankar$^1$, Leonardo Salvi$^2$, Maria Luisa Chiofalo$^{3, 1}$, Nicola Poli$^2$\footnote[7]{also at CNR-INO, e-mail: nicola.poli@unifi.it}\ and Murray J. Holland$^1$}
\address{$^1$ JILA, NIST, and Department of Physics, University of Colorado, 440 UCB, 
Boulder, CO  80309, USA}
\address{$^2$ Dipartimento  di  Fisica  e  Astronomia  and  LENS  -  Universit\`a  di  Firenze, INFN  -  Sezione  di  Firenze,  Via  Sansone  1,  50019  Sesto  Fiorentino,  Italy  }
\address{$^3$ Dipartimento di Fisica ``Enrico Fermi'',  Universit\`a di Pisa and INFN, Largo B. Pontecorvo 3,  I-56127 Pisa,  Italy}
\ead{\mailto{athreya.shankar@colorado.edu}}

\date{April 2019}

\begin{abstract}
    Bragg interferometers, operating using pseudospin-1/2 systems composed of two momentum states, have become a mature technology for precision measurements. State-of-the-art Bragg interferometers are rapidly surpassing technical limitations and are soon expected to operate near the projection noise limit set by uncorrelated atoms. Despite the use of large numbers of atoms, their operation is governed by single-atom physics. Motivated by recent proposals and demonstrations of Raman gravimeters in cavities, we propose a scheme to squeeze directly on momentum states for surpassing the projection noise limit in Bragg interferometers. In our modeling, we consider the unique issues that arise when a spin squeezing protocol is applied to momentum pseudospins. Specifically, we study the effects of the momentum width of the atomic cloud and the coupling to momentum states outside the pseudospin manifold, as these atoms interact via a mode of the cavity. We show that appreciable levels of spin squeezing can be demonstrated in suitable parameter regimes in spite of these complications. Using this setting, we show how beyond mean-field techniques developed for spin systems can be adapted to study the dynamics of momentum states of interacting atoms. Our scheme promises to be feasible using current technology and is experimentally attractive because it requires no additional setup beyond what will be  required to operate Bragg interferometers in cavities.
\end{abstract}

\maketitle

\section{Introduction}

Quantum metrology with atomic and atom-like platforms has greatly
benefited from the demonstration of squeezed spin states
\cite{Kitagawa1993,Ma2011PhyRep,giovannetti2004Sci,pezze2018RMP}
capable of overcoming the standard quantum limit (SQL) that arises for
measurement precision with uncorrelated pseudospins. Several of these
schemes rely on the availability of a common channel, such as a cavity
mode \cite{bohnet2014Nat} or a shared vibrational mode
\cite{bohnet2016Sci}, which couples to all the constituent
pseudospin-1/2 particles, thereby enabling entanglement generation via
collective quantum non-demolition measurements
\cite{kuzmich2000PRL,hosten2016Nature,bohnet2014Nat,cox2016PRL} or
deterministically via effective spin-spin interactions
\cite{bohnet2016Sci,leroux2010PRL,schleierSmith2010PRA}.

Bragg interferometers are widely used for applications such as tests
of fundamental physics and precision measurements of gravitational
acceleration
\cite{kovachy2015Nat,asenbaum2017PRL,yu2019AnnDePhyk,aguila2018NJP}. These
systems are attractive because of the unique encoding of the spin-1/2
system in two momentum states associated with the center-of-mass
motion of the atomic wavepacket. They operate by splitting the
wavepacket into two momentum states---that propagate along different
spatial paths accumulating a relative phase---and finally recombining
them to obtain interference fringes. Throughout the interferometer
operation, the atom is confined to the same metastable electronic
state, typically the ground state. Although an atom's momentum is a
continuous variable, a pseudospin-1/2 system with two discrete states
can be mapped on to the external motion in a Bragg
interferometer. This mapping requires an initial atomic momentum
distribution that is a sharp peak about a central value, so that the
distribution serves as one of the pseudospin states. Subsequently,
pulses of light resonantly couple this distribution to a second narrow
peak that is shifted by an even multiple of the well-defined photon
momentum. This shifted distribution serves as the other pseudospin
state.

Current Bragg interferometers
\cite{chiow2011PRL,jaffe2018PRL,hu2017PRL} operate in free space with
state-of-the-art technology enabling control of large numbers of
atoms. This technical progress has now achieved high signal-to-noise
ratios for determining the relative phase shift. However, despite the
use of large atom numbers, their operation can be completely described
in terms of single-atom physics since the atoms are uncorrelated. As a
result, regardless of whether further technical improvements are
realized, the phase sensitivity of these interferometers in the near
future will be fundamentally constrained by the SQL of
$\Delta\phi_\text{SQL} = 1/\sqrt{N}$ radians, where $N$ is the number
of atoms. Monotonically increasing $N$ to improve precision suffers
from problems such as practical limitations in trapping and cooling,
and uncontrolled phase changes that arise from atomic
collisions. Schemes to produce squeezed states of momentum pseudospins
are therefore attractive as a means to achieve precision beyond the
corresponding SQL for a given $N$. A major hurdle to producing such
states in Bragg interferometers is that squeezing requires a channel
for the atoms to controllably interact with each other, and such a
channel is unavailable in current interferometer designs.

The recent demonstration of a hyperfine-changing Raman gravimeter
operating inside an optical cavity \cite{hamilton2015PRL} motivates us
to envisage a similar operation of Bragg interferometers in cavities
in the near future. The availability of a cavity mode naturally opens
up a channel for mediating atom-atom interactions. Previous proposals
for cavity-based squeezing on momentum spins \cite{salvi2017PRL}
require significant experimental overhead dedicated to achieving
squeezing while the actual interferometer itself operates in free
space. In this work, we propose an alternative approach that marries
the generation of cavity-mediated spin squeezing
\cite{Ma2011PhyRep,hu2017PRA,norcia2018Sci} with the well known
advantages of operating the entire interferometer inside a cavity
\cite{hamilton2015PRL}. Importantly, our scheme does not require any
experimental overhead to generate interactions beyond what is already
needed to run a Bragg interferometer in a cavity. In fact, we show how
all-to-all atomic interactions are generated by simply switching off
one of the two Bragg lasers and suitably adjusting the frequency and
power of the other.

The use of momentum pseudospins in Bragg interferometers necessitates
two unique considerations. First, the atomic cloud will always have a
non-zero momentum width even after velocity selection. This width can
typically be neglected in the analysis with uncorrelated
atoms. Second, momentum pseudospins cannot be considered as closed two
level systems since the same pair of counterpropagating
electromagnetic fields couples the pseudospin states to other momentum
states, albeit with varying detunings. As a result, leakage to
undesirable momentum states is unavoidable even while applying
efficient Bragg pulses for spin rotations, and also when attempting to
engineer interactions for spin squeezing. In our work, we account for
the momentum width as well as leakage to undesirable states and show
that they can be important when considering the efficiency of a spin
squeezing protocol applied to momentum pseudospins. Nevertheless, as
we demonstrate, appreciable spin squeezing can still be achieved under
suitable and potentially realizable operating conditions.

In the process of accounting for the effects of momentum width and
losses to undesirable states, we show how to extend modeling
techniques originally developed for spin systems to interacting atoms
in matter-wave interferometers where information is encoded in
external degrees of freedom. This ability to map the continuous
momentum variable onto a discrete quantum pseudospace allows us to
directly employ methods developed for finite dimensional systems 
\cite{schachenmayer2015PRX, orioli2017PRA, lepoutre2019NatComm} . The
techniques we use to study our system are widely applicable for
investigations of beyond mean-field physics in a broad range of
schemes involving interacting atoms whose momentum states are coupled
by electromagnetic fields.

This article is organized as follows. In Section~\ref{sec:model}, we first derive a master equation describing the atom-cavity interactions. Further, we  adiabatically eliminate the cavity mode to arrive at an effective master equation for the atoms only. We also describe approximate numerical methods for each of the two master equations that enable us to obtain complementary insights into the squeezing dynamics. In Section~\ref{sec:squeezing}, we study the efficiency of squeezing on momentum pseudospins by considering Bragg transitions on the ${}^1\text{S}_0 - {}^3\text{P}_1$ transition in Strontium as a specific example \cite{aguila2018NJP}. We show that appreciable spin squeezing can be demonstrated using modest laser powers. We study the interplay of squeezing and superradiance, the dynamics under very fast squeezing, and the effect of a non-zero momentum width. We also discuss the manifestation of an experimentally observable many-body energy gap. In Section~\ref{sec:conc}, we conclude with comments on our results and possible extensions of our work. 

\section{\label{sec:model}Model and Methods}

\subsection{Setup}
\begin{figure}[!htb]
    \centering
    \includegraphics[width=\columnwidth]{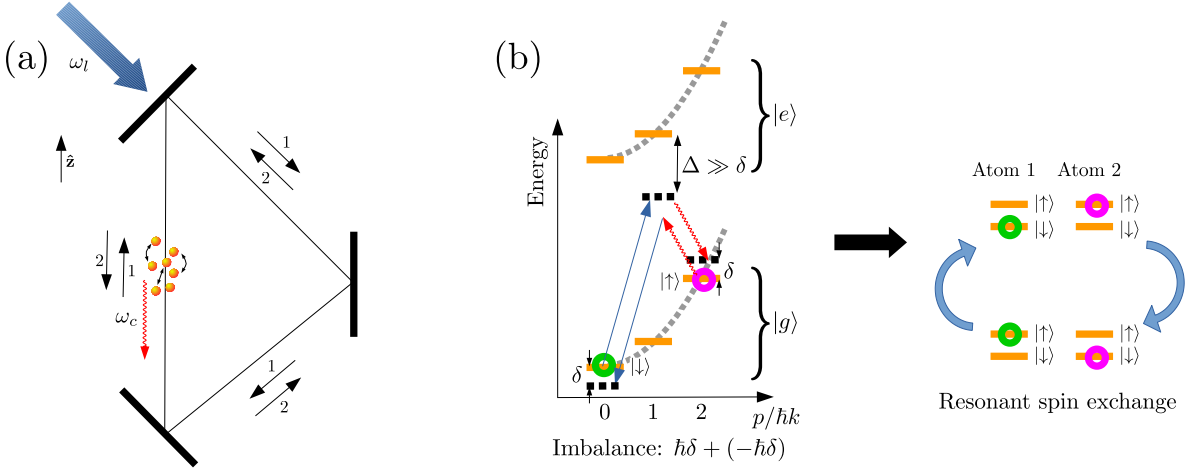}
    \caption{Experiment setup and working principle. (a) A cloud of atoms interacts with two counterpropagating modes of a ring cavity. One mode (mode 1) is driven at frequency $\omega_l$, while the counterpropagating mode (mode 2) is in vacuum, i.e. not pumped. The scheme enables cavity-mediated interactions between every pair of atoms. (b) The excitation or de-excitation of a single atom is  off-resonant. However, the exchange of excitation between two atoms is a resonant process.} 
    \label{fig:expt_setup}
\end{figure}

We consider a collection of $N$ atoms with mass $M$ in a ring cavity with resonance frequency $\omega_c$ as shown in Fig.~\ref{fig:expt_setup}(a). Each atom consists of two electronic levels $\ket{g}$ and $\ket{e}$ with transition frequency $\omega_a$. A laser with frequency $\omega_l$ drives one mode of the ring cavity. The cavity resonance is red detuned from the atomic transition by a detuning $\Delta_c = \omega_a - \omega_c > 0$, while the drive laser is detuned by $\Delta_l = \omega_a - \omega_l > 0$. The relative detuning of the laser and the cavity is $\Delta_{cl} = \omega_l - \omega_c \ll \Delta_c, \Delta_l$. Upon absorption or emission of a photon with wavevector $\mbf{k}$, the momentum of an atom is shifted by $\hbar k$, where $k = \abs{\mbf{k}}$.

\subsection{\label{sec:principle}Basic working principle}
The underlying principle for how squeezing is generated in our scheme is summarized in Fig.~\ref{fig:expt_setup}(b) for the case when the pseudospin is encoded in the states $\ket{\downarrow}\equiv\ket{g,0}$ and $\ket{\uparrow}\equiv\ket{g,2\hbar k}$ with transition frequency $4\omega_r$, where $\omega_r = \hbar k^2/2M$ is the atomic recoil frequency. We denote the driven mode as mode 1 and the counterpropagating mode as mode 2. The drive laser frequency is arranged such that $\Delta_{cl} = \omega_l-\omega_c = 4\omega_r+\delta$, where $\delta$ is a two-photon detuning typically assumed to be $>0$ in this work. The excitation of an atom from $\ket{\downarrow}$ to $\ket{\uparrow}$ (green circle in Fig.~\ref{fig:expt_setup}) is facilitated by the absorption of a drive photon and subsequent emission into mode 2. The energy imbalance between the photon exchange and the spin excitation is 
\begin{equation}
    \Delta E_{\downarrow\rightarrow\uparrow} = \left(\hbar\omega_l-\hbar\omega_c\right)-4\hbar\omega_r = \hbar\delta. 
    \label{eqn:delta_e1}
\end{equation}
Similarly, the de-excitation of a second spin (magenta circle in Fig.~\ref{fig:expt_setup}) is accompanied by the absorption of a photon in mode 2 and subsequent emission at the drive frequency, leading to an energy imbalance 
\begin{equation}
    \Delta E_{\uparrow\rightarrow\downarrow} = \left(\hbar\omega_c-\hbar\omega_l\right)-(-4\hbar\omega_r) = -\hbar\delta. 
    \label{eqn:delta_e2}
\end{equation}
However, from Eqs~(\ref{eqn:delta_e1}) and (\ref{eqn:delta_e2}), the simultaneous excitation of one atom and de-excitation of the other is resonant, facilitated by the four-photon process consisting of absorption of a drive photon, emission and absorption of a virtual cavity photon and subsequent return of the photon to the drive laser. Assuming that the cavity mode couples identically to all the atoms, the cavity mode cannot distinguish which atom was excited and which one was de-excited, leading to an effective Hamiltonian of the form 
\begin{equation}
    \hat{H}_\text{eff} \propto \hat{J}^-\hat{J}^+,
    \label{eqn:basic}
\end{equation}
\noindent where $\hat{J}^\pm = \sum_{j=1}^N \hat{\sigma}_j^\pm$, with $\hat{\sigma}_j^+ = \ket{\uparrow}_j\bra{\downarrow}$ and $\hat{\sigma}_j^- = \left(\hat{\sigma}_j^+\right)^\dag$. The effective Hamiltonian, Eq.~(\ref{eqn:basic}), can be expressed as 
\begin{equation}
    \hat{J}^-\hat{J}^+ = \hat{\mathbf{J}}\cdot\hat{\mathbf{J}} - \hat{J}^z \hat{J}^z - \hat{J}^z,
    \label{eqn:jmjp_breakdown}
\end{equation}
where $\hat{\mathbf{J}} = \hat{J}^x \mathbf{\hat{x}} + \hat{J}^y \mathbf{\hat{y}} + \hat{J}^z \mathbf{\hat{z}}$ and $\hat{J}^i$ for $i=x,y,z$ are the Cartesian components of the collective angular momentum $\hat{\mathbf{J}}$ formed by the momentum pseudospins. The second term is the familiar one-axis twisting (OAT) interaction that gives rise to spin squeezed states useful for quantum metrology \cite{Ma2011PhyRep}. The first term, on the other hand, opens a many-body energy gap that has been experimentally observed, for example, using spins encoded in optical clock transitions \cite{norcia2018Sci}. We briefly discuss how the latter effect manifests in our system in Section~\ref{sec:gap_protect}.

\subsection{Atom-cavity interactions}

We now proceed to derive a master equation that reflects the underlying atom-cavity interaction at the heart of the resonant spin exchange intuitively described in the previous section. The Hamiltonian governing the dynamics of the atom-cavity system is 
\begin{eqnarray}
\hat{H} &=& \sum_{j=1}^N \left( \frac{\hat{p}_j^2}{2M} + \frac{\hbar \omega_a}{2}(\ket{e}_j\bra{e}-\ket{g}_j\bra{g}) \right) + \sum_{s=1}^2 \hbar \omega_c \hat{a}_s^\dag \hat{a}_s \nonumber\\
&+& \sum_{j=1}^N \sum_{s=1}^2 \frac{\hbar g}{2}\left( \hat{a}_s e^{i k_s \hat{z}_j}\ket{e}_j\bra{g} + \hat{a}_s^\dag e^{-i k_s \hat{z}_j}\ket{g}_j\bra{e}  \right) \nonumber\\
&+& \hbar \sqrt{\kappa} \left( \alpha  e^{-i \omega_l t}\hat{a}_1^\dag + \alpha^* e^{i \omega_l t} \hat{a}_1\right),
\end{eqnarray}
where $g$ is the atom-cavity vacuum Rabi frequency, $\kappa$ is the cavity decay rate and $\alpha$ is the amplitude of the drive laser with $\abs{\alpha}^2$ the photon flux in units of photons/time. The operators $\hat{a}^\dag_s,\hat{a}_s$ for $s=1,2$ respectively describe the creation and annihilation of photons in modes 1 and 2 whose wavevectors satisfy $\mbf{k}_1 = -\mbf{k}_2 = k \mbf{\hat{z}}$ (see Fig.~(\ref{fig:expt_setup})(a)). The operator $\hat{z}_j$ represents the position of atom $j$ along the cavity axis. The decay of the cavity fields is accounted for by the standard Lindblad dissipator of the type $\mD[\hat{O}]\rho=\hat{O}\rho \hat{O}^\dag - \hat{O}^\dag \hat{O} \rho/2 - \rho \hat{O}^\dag\hat{O}/2$ for jump operator $\hat{O}$ and density matrix $\rho$. The resulting master equation is 
\begin{equation}
    \dot{\rho} = \frac{1}{i\hbar}\left[ \hat{H}, \rho \right] + \sum_{s=1}^2 \kappa \mD[\hat{a}_s]\rho.
    \label{eqn:master_eqn}
\end{equation}
We neglect free-space scattering in our analysis since superradiant decay (Section~(\ref{sec:superrad})) is typically the dominant dissipation mechanism (see \ref{sec:fss} and discussion in Section~\ref{sec:conc}).

We work in an interaction picture rotating at the drive frequency $\omega_l$ with free evolution Hamiltonian $H_f = \sum_j \hbar \omega_l/2(\ket{e}_j\bra{e}-\ket{g}_j\bra{g}) + \sum_{s} \omega_l \hat{a}_s^\dag \hat{a}_s$. First, we adiabatically eliminate the excited state $\ket{e}$ based on the large detuning of the drive lasers and cavity modes from the $\ket{g}\leftrightarrow\ket{e}$ transition (\ref{sec:ad_exc}). Further, on long timescales, the upwards propagating mode (mode 1) is composed of a macroscopic steady state amplitude with small fluctuations around this value. The macroscopic amplitude $\beta$ ($\abs{\beta}\gg 1$) is found from the mean-field equation \begin{equation}
    \dot{\beta} = -\left( \frac{\kappa}{2}-i\Delta_{cl}\right)\beta -i\sqrt{\kappa} \alpha.
\end{equation}
For $t\gg \kappa^{-1}$, the steady-state value is $\beta = -i\sqrt{\kappa} \alpha/(\kappa/2-i\Delta_{cl})$. We displace mode 1 by the amplitude $\beta$ by making the transformation $\hat{a}_1 \rightarrow \beta + \hat{a}_1$. Apart from introducing some constant terms that can be neglected, the resulting Hamiltonian is  
\begin{eqnarray}
\hat{H}_{a-c} &=& \sum_{j=1}^N \frac{\hat{p}_j^2}{2M}  - \sum_{s=1}^2 \hbar \Delta_{cl} \hat{a}_s^\dag \hat{a}_s \nonumber\\
&-& \sum_{j=1}^N \frac{\hbar g^2}{4\Delta_c}\left( \beta^* \hat{a}_2 e^{-i k_\text{eff} \hat{z}_j} + \beta \hat{a}_2^\dag e^{i k_\text{eff} \hat{z}_j}\right),
\label{eqn:ham_atom_cav}
\end{eqnarray}
with $k_\text{eff}=k_1-k_2=2k$ the effective wavevector along $\mbf{\hat{z}}$. The dissipative part of the master equation remains the same. The second line of Eq.~(\ref{eqn:ham_atom_cav}) reflects the dominant photon exchange between the macroscopic field in mode 1 and the vacuum of mode 2. In writing Eq.~(\ref{eqn:ham_atom_cav}), we have neglected the small exchange process between the vacuum fields of the two modes. This approximation allows us to keep track of only mode 2 and ignore the other terms containing $\hat{a}_1$ in the master equation since mode 1 only interacts with the atoms and mode 2 through the $c$-number $\beta$.

\subsubsection{Momentum width using the $\ket{n,q}$ notation}

The momentum shift operator $e^{\pm i \kE \hat{z}_j}$ appearing in Eq.~(\ref{eqn:ham_atom_cav}) can only shift the momentum in units of $\hbar\kE$. For simplicity, we consider initial atomic states that are clustered around $\ket{\downarrow}\equiv\ket{\nDown\hbar\kE}$, i.e. $\ev{\hat{p}_j(0)} = \nDown\hbar\kE$, where $\nDown$ is an integer. (A superposition of $\ket{\downarrow}$ and $\ket{\uparrow}\equiv \nUp\hbar\kE$ with $\nUp=\nDown+1$ can be subsequently obtained by a $\pi/2$ Bragg pulse.) We introduce two labels $n,q$ to represent a momentum state as $\ket{p}\equiv\ket{n,q}$. The label $n$ denotes the momentum center and is defined as 
\begin{equation}
    n = \nif{\frac{p}{\hbar\kE}}
    \label{eqn:n_def}
\end{equation}
where $\nif{x}$ denotes the nearest integer to $x$. The label $q$ quantifies the deviation from a center and is defined as 
\begin{equation}
    q = p - n \hbar\kE.
    \label{eqn:q_def}
\end{equation}
We note that an initial offset from an integer multiple of $\hbar\kE$ , i.e. $\ev{\hat{p}_j(0)} = \nDown\hbar\kE + p_\text{off}$ can be trivially accounted for by denoting states as $\ket{p}\equiv\ket{n,q,p_\text{off}}$ so that $p=n\hbar\kE+q+p_\text{off}$. Note that $\abs{p_\text{off}}<1/2$ since larger values can be modeled as an offset about a shifted initial center $\nDown\rightarrow\nDown \pm 1$. Without loss of generality, we assume $p_\text{off}=0$.

The momentum width is characterized by a spread $\sigma_q$. We assume that the initial momentum spread $\sigma_q$ is small compared to the difference between subsequent centers, i.e. 
\begin{equation}
    \frac{\sigma_q}{\hbar\kE} \ll 1.
    \label{eqn:sigma_q_req1}
\end{equation}
Combined with the fact that the dynamics under Eq.~(\ref{eqn:ham_atom_cav}) does not change the spread but only shifts the center, this assumption ensures that we can assume the orthogonality relation 
\begin{equation}
\langle n',q' \lvert n,q\rangle = \delta_{n,n'}\delta(q-q').
\label{eqn:ortho}
\end{equation}

\subsection{Numerical solution: Semiclassical Langevin equations}

The master equation governing the atom-cavity interactions is:
\begin{equation}
    \dot{\rho}_{a-c} = \frac{1}{i\hbar}\left[ \hat{H}_{a-c}, \rho_{a-c} \right] + \kappa \mD[\hat{a}_2]\rho.
    \label{eqn:master_eqn_dtwa}
\end{equation}

\noindent The momentum shift operator $e^{i \kE \hat{z}_j}$ can be expressed as 
\begin{equation}
 e^{i \kE \hat{z}_j} = \sum_{n=-\infty}^{\infty} \ket{n+1,q}_j\bra{n,q}. 
 \label{eqn:mom_shift_decomp_2}
\end{equation}
We define generalized population and coherence operators $\hSig_{nm}^j$ as 
\begin{equation}
    \hSig_{nm}^j = \ket{n,q_j}_j\bra{m,q_j}.
    \label{eqn:sig_def}
\end{equation}
We have dropped the label $q_j$ in defining $\hSig_{nm}^j$, since $q_j$ does not change under the dynamics governed by the master equation. The free energy term can be expressed as 
\begin{equation}
     \frac{\hat{p}_j^2}{2M} = \sum_{n=-\infty}^{\infty}\hbar \omega_{n}^j \hSig_{nn}^j,
\end{equation}
where $\hbar\omega_n^j = (n\hbar\kE+q_j)^2/2M$. The frequency $\omega_n^j$ can be better expressed as 
\begin{equation}
    \omega_n^j = 4\omega_r\left(n^2 + 2n\tdq_j\td{\sigma}_q + \tdq_j^2 \td{\sigma}_q^2 \right),
    \label{eqn:omega_nj}
\end{equation}
where we have introduced the dimensionless quantities $\tdq = q/\sigma_q$ and $\td{\sigma}_q = \sigma_q/\hbar\kE$. The Hamiltonian, Eq.~(\ref{eqn:ham_atom_cav}), can now be expressed as 
\begin{eqnarray}
\hat{H} = \sum_j \sum_n\hbar \omega_{n}^j \hSig_{nn}^j  - \hbar \Delta_{cl} \hat{a}_2^\dag \hat{a}_2
- \sum_j \sum_n \frac{\hbar g^2}{4\Delta_c}\left( \beta^* \hat{a}_2 \hSig_{n,n+1}^j + \beta \hat{a}_2^\dag \hSig_{n+1,n}^j \right). \nonumber\\
\label{eqn:ham_dtwa_2}
\end{eqnarray}
Expressed this way, the atom-cavity interaction is reminiscent of the detuned Tavis-Cummings model that is at the heart of cavity-based spin exchange schemes considered for optical clock transitions \cite{hu2017PRA,norcia2018Sci} . 

We write down the dynamical equations for the corresponding $c$-numbers $s_{nm}^j \leftrightarrow \hSig_{nm}^j$ and $\zeta \leftrightarrow \hat{a}_2$ within the truncated Wigner approximation (TWA) framework \cite{orioli2017PRA}. We introduce the effective coupling strength $\gE = g^2 \abs{\beta}/2\Delta_c$ and without loss of generality assume that $\beta$ is real. These equations are 
\begin{eqnarray}
    &&\frac{d}{dt} s_{nm}^j = -i(\omega_m^j-\omega_n^j) s_{nm}^j + i\frac{\gE}{2} \left( 
    \zeta^*\left(s_{n,m-1}^j-s_{n+1,m}^j\right)
    +\zeta\left(s_{n,m+1}^j - s_{n-1,m}^j\right)\right), \nonumber\\
    &&\frac{d}{dt}\zeta = -\left(\frac{\kappa}{2}-i\Delta_{cl}\right)\zeta + i\frac{\gE}{2}\sum_j\sum_n s_{n+1,n}^j + \sqrt{\frac{\kappa}{4}}\left( \xi_1(t) + i \xi_2(t) \right),
    \label{eqn:dtwa_mf}
\end{eqnarray}
where $\xi_l(t)$ for $l=1,2$ are white-noise processes satisfying $\overline{\xi_l(t)}=0$ and $\overline{\xi_l(t)\xi_{l'}(t')} = \delta_{l,l'}\delta(t-t')$. The bar indicates averaging over several trajectories with different noise realizations (and initial conditions, see Section~\ref{sec:mcm_init_con}) .
The equation for $\zeta$ is a stochastic differential equation because of the noise arising from coupling to  modes outside the cavity \cite{tieri2017arXiv}. We refer to this model as the multi-center model (MCM) because of its ability to track an arbitrary number of momentum centers.

\subsubsection{\label{sec:mcm_init_con}Initial Conditions}
The mode amplitude $\zeta = a+ib$ is initialized according to the Wigner distribution of a vacuum state as 
\begin{equation}
    P(a,t=0) = P(b,t=0) = \mathcal{N}(0,1/2),
    \label{eqn:mode_wgn}
\end{equation}
so that $\overline{\abs{\zeta}^2} = \left(\ev{\hat{a}_2^\dag\hat{a}_2} + \ev{\hat{a}_2\hat{a}_2^\dag} \right)/2 = 1/2$. Here, $\mathcal{N}(\mu,\sigma)$ denotes a normal distribution with mean $\mu$ and standard deviation $\sigma$.

For the atoms, we first consider each atom to be in a state described by the density matrix 
\begin{equation}
    \rho^{(1)}(0) = \frac{1}{\sqrt{2\pi}\sigma_q} \int_{-\infty}^{\infty} dq \; e^{-q^2/2 \sigma_q^2} \ket{\nDown,q} \bra{\nDown,q}
    \label{eqn:init_state}
\end{equation}
where the restriction Eq.~(\ref{eqn:sigma_q_req1}) ensures that states with $\abs{q}\sim\hbar\kE$ do not contribute significantly so that the limits of integration can be extended to $\pm\infty$. 

We note that by using two labels $n,q$ to characterize the momentum, we have effectively split the momentum phase space distribution into one for the discrete label $n$ and one for the continuous label $q$. From Eq.~(\ref{eqn:omega_nj}), the effect of $q_j$ is to modify the frequencies $\omega_n^j,\omega_m^j,\ldots$ for each $j$. A state described by Eq.~(\ref{eqn:init_state}) can be simulated by assuming that in each trajectory, the value of $q_j$ for each atom is drawn according to a normal distribution characterized by a spread $\sigma_q$, as 
\begin{equation}
    P(q_j,t=0) = \mathcal{N}(0,\sigma_q) \implies P(\tdq_j,t=0) = \mathcal{N}(0,1) .
\end{equation}

To appropriately sample the $n$-space distribution corresponding to the state described by Eq.~(\ref{eqn:init_state}), we note that the discrete levels $n,m,\ldots$ are reminiscent of the different $m_J$ levels in a $2J+1$ spin manifold. Here, the choice of $J$ depends on the number of discrete levels that participate significantly in the dynamics. We initialize the $c$-numbers $s_{nm}^j$ according to the DTWA (discrete truncated Wigner approximation) prescription \cite{schachenmayer2015PRX, orioli2017PRA}, namely,
\begin{eqnarray}
    &&s_{\nDown,\nDown}^j = 1, \nonumber\\
    &&P(2\;\text{Re}\{s_{\nDown,m\neq\nDown}^j\} = \pm 1) = P(2\;\text{Im}\{s_{\nDown,m\neq\nDown}^j\} = \pm 1) = \frac{1}{2}, \nonumber\\
    &&s_{m\neq\nDown,\nDown}^j = \left(s_{\nDown,m\neq \nDown}^j\right)^*, \nonumber\\
    &&s_{n \neq \nDown, m \neq \nDown}^j = 0.
    \label{eqn:mcm_init}
\end{eqnarray}
We note that our choice of initial conditions is consistent with a formal generalization of the Truncated Wigner Approximation technique to systems with $D$ discrete states on a given site \cite{lepoutre2019NatComm}. For generating squeezing in our model, we require each atom to be initialized in an equal superposition of the $\nDown,\nUp$ centers. Starting with the initial conditions in Eq.~(\ref{eqn:mcm_init}), we obtain the $c$-number values corresponding to such an equal superposition by implementing a fictitious instantaneous state rotation that rotates each spin to lie on the equatorial plane of the Bloch sphere formed by $\nDown,\nUp$ (\ref{sec:mcm_rot}). The observables from the MCM simulations are averaged over $2000$ trajectories in order to sample the initial conditions and noise realizations.

\subsection{\label{sec:atom-atom}Effective atom-atom interactions}

The spin exchange dynamics anticipated in Section~\ref{sec:principle} is confirmed when mode 2 is adiabatically eliminated to obtain a master equation describing the effective atom-atom interactions.
When mode 2 is negligibly excited, it can be considered as a reservoir in a vacuum state with density matrix $R_0 = \ket{0}\bra{0}$. We use the superoperator formalism to adiabatically eliminate mode 2 \cite{carmichael2010statistical}. The details of this derivation are presented in \ref{sec:mode2_elim}. The resulting master equation can be compactly expressed in terms of operators analogous to collective angular momentum operators, which we introduce below.

\subsubsection{\label{sec:coll_ang}Collective angular momentum operators}

First, we introduce generalized population and coherence operators for a single atom, along the lines of Eq.~(\ref{eqn:sig_def}), but with an extra label $q$, as \begin{equation}
    \hSig_{nm}^{j,q} = \ket{n,q}_j\bra{m,q}.
    \label{eqn:sig_def_q}
\end{equation}
We can then define collective angular momentum operators $\hat{\mJ}_n^\pm, \hat{\mJ}_n^z$ acting on any two consecutive momentum centers $n,n+1$ as 
\begin{equation}
    \hat{\mJ}_n^+ = \sum_{j=1}^N\int_{-\infty}^{\infty}dq\;\hat{\sigma}_{n+1,n}^{j,q}, \;
    \hat{\mJ}_n^- = \sum_{j=1}^N\int_{-\infty}^{\infty}dq\;\hat{\sigma}_{n,n+1}^{j,q}, \;
    \hat{\mJ}_n^z = \frac{1}{2}\sum_{j=1}^N\int_{-\infty}^{\infty}dq\;\left(\hat{\sigma}_{n+1,n+1}^{j,q} - \hat{\sigma}_{n,n}^{j,q}\right).
\end{equation}

\noindent With $\hat{\mJ}_n^x = \left(\hat{\mJ}_n^+ + \hat{\mJ}_n^-\right)/2$ and $\hat{\mJ}_n^y = \left(\hat{\mJ}_n^+ - \hat{\mJ}_n^-\right)/2i$, the operators $\hat{\mJ}_n^x,\hat{\mJ}_n^y,\hat{\mJ}_n^z$ satisfy the usual angular momentum commutation relations 
\begin{equation}
    \left[\hat{\mJ}_n^j, \hat{\mJ}_n^k \right] = i\epsilon_{jkl}\hat{\mJ}_n^l,
    \label{eqn:ang_comm}
\end{equation}

\noindent where $\epsilon_{jkl}$ is the usual Levi-Civita symbol for the right-handed coordinate system formed by the $x,y,z$ axes. Once again, the restriction on initial states, Eq.~(\ref{eqn:sigma_q_req1}), ensures that the limits of integration over $q$ can be extended to $\pm\infty$ while still allowing the use of the orthogonality relation Eq.~(\ref{eqn:ortho}) in deriving the commutation rules in Eq.~(\ref{eqn:ang_comm}). Specifically, the collective spin consisting of the pseudospin-1/2 systems formed by the two centers $\nDown,\nUp$ are characterized by the operators $\hat{\mJ}_{\nDown}^\pm, \hat{\mJ}_{\nDown}^z$.  

The master equation for the reduced density matrix $\rho_a=\text{Tr}_{c}\left[\rho_{a-c}\right]$, obtained after adiabatically eliminating mode 2, can be expressed as 
\begin{equation}
      \dot{\td{\rho}}_a= \frac{1}{i\hbar}\left[\hat{H}_\text{eff}, \td{\rho}_a \right] + \sum_{n=-\infty}^{\infty} 2 \Gamma_n \lv[\hat{\mJ}_n^+]\td{\rho}_a,
      \label{eqn:me_compact}
\end{equation}
with the effective Hamiltonian 
\begin{eqnarray}
    \hat{H}_\text{eff} = \sum_{j=1}^N \sum_{n=-\infty}^{\infty} \int_{-\infty}^{\infty}dq\; \left(8n\hbar\omega_r\right)\left(\tdq \td{\sigma}_q\right)\hat{\sigma}_{n,n}^{j,q}
    + \sum_{n=-\infty}^{\infty}\hbar\chi_n \hat{\mJ}_n^- \hat{\mJ}_n^+,
    \label{eqn:h_eff_2}
\end{eqnarray}
where the coherent and dissipative coupling strengths, $\chi_n$ and $\Gamma_n$ are defined as 
\begin{eqnarray}
    \chi_n = \left(\frac{g^2\abs{\beta}}{4\Delta_c}\right)^2 \frac{\delta_n}{\kappa^2/4+\delta_n^2}, \;
    \Gamma_n = \left(\frac{g^2\abs{\beta}}{4\Delta_c}\right)^2 \frac{\kappa/2}{\kappa^2/4+\delta_n^2},\nonumber\\
    \label{eqn:coup_strength_main}
\end{eqnarray}
with $\delta_n \equiv \Delta_{cl}-4\omega_r\left(1+2n\right)$.
The notation $\td{\rho}_a$ indicates that the master equation is written in an appropriate interaction picture (\ref{sec:mode2_elim}).

\subsection{Numerical solution: Cumulant theory for one and two-atom operators}

To make computations tractable, we assume that the $\nDown,\nUp$ centers form a closed two-level system while studying the collective spin dynamics using the master equation, Eq.~(\ref{eqn:me_compact}). To this effect, we truncate the master equation, Eq.~(\ref{eqn:me_compact}), as 
\begin{equation}
      \dot{\td{\rho}}_a= \frac{1}{i\hbar}\left[\hat{H}_\text{eff}^{\text{T}}, \td{\rho}_a \right] + 2 \Gamma_{\nDown} \lv[\hat{\mJ}_{\nDown}^+]\td{\rho}_a,
      \label{eqn:me_truncated}
\end{equation}
where the truncated Hamiltonian is 
\begin{eqnarray}
    \hat{H}_\text{eff}^{\text{T}} = \sum_{j=1}^N \sum_{n=\nDown,\nUp} \int_{-\infty}^{\infty}dq\; \left(8n\hbar\omega_r\right)\left(\tdq\td{\sigma}_q\right)\hat{\sigma}_{n,n}^{j,q}
    + \hbar\chi_{\nDown} \hat{\mJ}_{\nDown}^- \hat{\mJ}_{\nDown}^+.
    \label{eqn:h_eff_2_truncated}
\end{eqnarray}
We recall that the effective Hamiltonian, Eq.~(\ref{eqn:h_eff_2_truncated}), with $\td{\sigma}_q=0$ is analogous to the standard spin exchange/one-axis twisting model studied for closed two-level systems coupled to a cavity \cite{Ma2011PhyRep,hu2017PRA,norcia2018Sci,swan2018PRL} (also compare with Eq.~(\ref{eqn:basic})), and provides a reference model against which complications arising from the nature of momentum states can be contrasted. 

Exact solutions even for the truncated master equation, Eq.~(\ref{eqn:me_truncated}), are computationally intractable because of the exponential scaling of the Liouville space with atom number. We use an approximate method where we only keep track of expectation values of single atom and two atom operators, of the type $\ev{\hat{\sigma}_{n_a,n_b}^{1,q}}$ and $\ev{\hat{\sigma}_{n_a,n_b}^{1,q}\hat{\sigma}_{n_c,n_d}^{2,q'}}$, where the $n$ values can take either $\nDown$ or $\nUp$. Since we are ignoring the other momentum centers, we refer to this model as the two-center model (TCM). As in the MCM, the single atom and two atom expectation values are first initialized according to the state described by Eq.~(\ref{eqn:init_state}). Next, an instantaneous rotation transforms these quantities to correspond to a state that is an equal superposition of $\nDown,\nUp$ (\ref{sec:tcm_rot}). The identical initial conditions for each atom and the permutation symmetry of the master equation enable us to avoid separate indices for every atom in the system, with the number of atoms $N$ explicitly appearing in the equations for the quantities $\ev{\hat{\sigma}_{n_a,n_b}^{1,q}}$ and $\ev{\hat{\sigma}_{n_a,n_b}^{1,q}\hat{\sigma}_{n_c,n_d}^{2,q'}}$. The resulting equations of motion for these single atom and two atom expectation values are summarized in \ref{sec:eom_cumulant}.

\section{\label{sec:squeezing}Spin squeezing}

\subsection{Figure of merit: Wineland squeezing parameter $\xi_\text{R}^2$}

The $\hat{\mJ}_{\nDown}^z \hat{\mJ}_{\nDown}^z$ term implicit in Eq.~(\ref{eqn:h_eff_2_truncated}) (see Eq.~(\ref{eqn:jmjp_breakdown})) can be exploited to prepare spin squeezed states. The figure of merit of a squeezed state, relevant for quantum metrology, is the Wineland squeezing parameter $\xi_\text{R}^2$ \cite{Ma2011PhyRep} defined as 
\begin{equation}
    \xi_\text{R}^2 = \frac{1}{\mathcal{C}^2} \frac{V_\text{min}}{V_\text{SQL}}.
    \label{eqn:wineland}
\end{equation}
The contrast $\mathcal{C}$ is given by 
\begin{equation}
    \mathcal{C} = \frac{\abs{\ev{\mbf{\hat{J}_{\nDown}}}}}{N/2},
    \label{eqn:ctrst}
\end{equation}
where $\mbf{\hat{J}_{\nDown}} = \hat{\mJ}_{\nDown}^x \mbf{\hat{x}} + \hat{\mJ}_{\nDown}^y \mbf{\hat{y}} + \hat{\mJ}_{\nDown}^z \mbf{\hat{z}}$. 
For a given state, $V_\text{min}$ is the variance in a spin component in the plane perpendicular to the mean spin direction (MSD), minimized over all axes in this plane. Mathematically,
\begin{equation}
    V_\text{min} = \min_{\mbf{\hat{n}} \perp \mbf{\hat{n}}_{\text{MSD}}} \left\langle\left(\mbf{\hat{J}_{\nDown}}\cdot \mbf{\hat{n}}\right)^2\right\rangle.
    \label{eqn:vmin}
\end{equation}

\noindent $V_\text{SQL}=N/4$ sets the corresponding SQL for unentangled atoms and is the variance of any spin component in this plane for a coherent spin state \cite{Ma2011PhyRep}. 

\subsection{Considerations for choosing parameters}

First, we note that the single atom-cavity vacuum Rabi frequency can be expressed as $g=\sqrt{C\kappa\gamma}$, where $C$ is the cooperativity of the cavity and $\gamma$ is the inverse lifetime of the excited state. Our model imposes two constraints that limit $\abs{\beta}$ to the range
\begin{equation}
    1 \ll \abs{\beta} \ll \frac{\Delta_c}{\sqrt{C\kappa\gamma}},
    \label{eqn:beta_constraint}
\end{equation}
where we have used $g=\sqrt{C\kappa\gamma}$. The lower bound $\abs{\beta}\gg 1$ allows us to treat mode 1 as a classical field represented by the $c$-number $\beta$. The upper bound ensures that the excited state $\ket{e}$ is negligibly populated, i.e. 
\begin{equation}
   \frac{g^2\abs{\beta}^2}{4\Delta_c^2}\ll 1,
\end{equation}
thereby ensuring that the adiabatic elimination of $\ket{e}$ is valid. We work with $\abs{\beta}$ values such that $\abs{\beta} \geq 100$ and the excited state population is $\leq 0.01$. 

\subsection{Parameters for the ${}^{1}\text{S}_0 - {}^{3}\text{P}_1$ transition in ${}^{88}\text{Sr}$}

Although our scheme is applicable to a wide variety of atomic species, here we consider its efficiency when it is implemented on the $689$ nm ${}^{1}\text{S}_0 - {}^{3}\text{P}_1$ transition of ${}^{88}\text{Sr}$. Our choice is motivated by the advantages of using ground-state  ${}^{88}\text{Sr}$ in Bragg interferometers \cite{aguila2018NJP}, such as its extremely small scattering cross-section, insensitivity to stray magnetic fields and ease of experimental manipulation, including accessing the parameter regimes required for our scheme. The inverse lifetime of the excited state is $\gamma/2\pi = 7.6$ kHz while the single photon recoil frequency is $\omega_r/2\pi=4.74$ kHz. The spin-1/2 system is encoded in $\ket{\downarrow}\equiv \ket{{}^{1}\text{S}_0,0 \hbar k}$ and $\ket{\uparrow}\equiv \ket{{}^{1}\text{S}_0,2 \hbar k}$ implying that $\nDown=0,\nUp=1$. We consider $N=10^3$ atoms in a cavity with decay rate $\kappa/2\pi=100$ kHz, and with either of two cooperativities, $C=1$ ($C=10$). The single atom-cavity vacuum Rabi frequency $g = \sqrt{C\kappa\gamma}$ then takes the value $g/2\pi \approx 27.6$ kHz ($87.2$ kHz). We assume that the cavity resonance is detuned from the atomic transition such that $\Delta_c/2\pi = 200$ MHz. We characterize the relative strength of the dissipative and dispersive interactions by the ratio $R$ defined as 
\begin{equation}
    R = \frac{\Gamma_{\nDown}}{\chi_{\nDown}} =  \frac{\kappa}{2\delta_{\nDown}}.
\end{equation}
Squeezing by one-axis twisting occurs when the dispersive interactions dominate, corresponding to the regime $R\ll1$. We consider $R$ in the range $0.025-0.2$ in our study. 
Model-enforced constraints (see Eq.~(\ref{eqn:beta_constraint}) and the discussion following it) restrict the photon number in mode 1, $\abs{\beta}^2$, to the range $1\times10^4 - 2 \times 10^6$ ($ 1\times10^4 - 2\times 10^5$). Experimentally, these constraints translate to varying the power $P$ in the drive laser in a range $10 \; \text{nW} - 150 \; \mu\text{W}$ ($10 \; \text{nW} - 15 \; \mu\text{W}$) (\ref{sec:param_cons}). In standard one-axis twisting with closed two level systems, squeezing proceeds at a characteristic rate $Q=N\chi_{\nDown}$ \cite{hu2017PRA,swan2018PRL,Ma2011PhyRep}. The permissible values of $\abs{\beta}^2$ results in a squeezing rate $Q/2\pi$ in the range $5 \; \text{Hz} - 7.6 \; \text{kHz}$ ($0.5 \; \text{kHz} - 76 \; \text{kHz}$) (\ref{sec:param_cons}). We only consider squeezing rates such that $Q/\delta_{\nDown} \ll 1$ ($<1/50$ in all simulations), allowing for the adiabatic elimination of mode 2 in deriving the two-center model (see \ref{sec:markov_validity}). Even in this regime, while very slow rates are undesirable from a technical perspective, very fast squeezing with $Q \gtrsim \omega_r$ leads to coupling with momentum states outside the pseudospin manifold and degrades the squeezing, as we will demonstrate. 

Finally, to account for the momentum width of the atomic cloud, we consider values $\td{\sigma}_q \leq 0.1$ to satisfy the requirement, Eq.~(\ref{eqn:sigma_q_req1}), of our model. The dephasing rate $\mu_d = 4\sqrt{2}\omega_r \td{\sigma}_q$ associates a characteristic timescale to the momentum width. Specifically, for a collection of atoms initialized in the same, equal superposition between the two centers $\nDown,\nUp$ and undergoing free evolution, the contrast $\mathcal{C}$ decays as $\mathcal{C}(t) = e^{-\mu_d^2 t^2}$. With $\td{\sigma}_q \leq 0.1$, the corresponding maximum rate is $\mu_d/2\pi=2.7 \; \text{kHz}$.

\subsection{\label{sec:superrad}Limits set by superradiance}

\begin{figure}[!htb]
    \centering
    \includegraphics[width=\columnwidth]{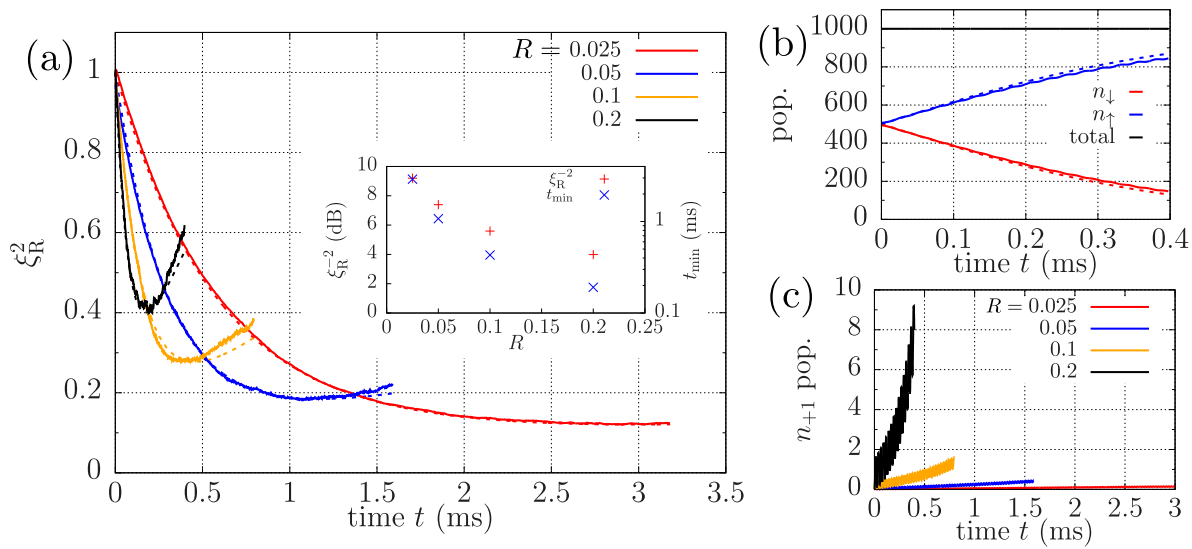}
    \caption{Interplay of squeezing and superradiance for different $R=\kappa/2\delta_{\nDown}$ values. (a) Evolution of $\xi_\text{R}^2$ for $R=0.025,0.05,0.1,0.2$. Inset: Maximum metrological gain (in dB) and time taken to achieve this gain. (b) Population in $\nDown$, $\nUp$ for $R=0.2$, with total population in all centers adding up to $N=10^3$. (c) Population in $n_{+1}$ for different $R$ values. In this panel, $N=10^3$, $C=1$, $\abs{\beta}^2 \approx 5.4 \times 10^5$. Solid (dashed) lines represent MCM (TCM) results. Four centers, $\nDown,\nUp,n_{+1},n_{+2}$, were tracked in the MCM simulations, with negligible population in $n_{+2}$.} 
    \label{fig:diffR_sameBeta}
\end{figure}

We first consider the case of $\td{\sigma}_q\approx 0$, i.e. negligible momentum width. Figure~\ref{fig:diffR_sameBeta}(a) plots the evolution of the spin squeezing parameter in the $C=1$ case for values of $R$ in the range $0.025-0.2$, and with $\abs{\beta}^2\approx 5.4 \times 10^5$. Modest laser powers, up to $40 \; \mu$W, are sufficient to maintain this intracavity photon number for the range of $R$ considered here (\ref{sec:param_cons}). In this parameter regime, the TCM (dashed) and MCM (solid) results agree excellently until $\xi_\text{R}^2$ reaches its minimum value. The minimum value of $\xi_\text{R}^2$ arises as a trade-off between the twisting dynamics that decreases $V_\text{min}$ (Eq.~(\ref{eqn:wineland}) and fluctuations in superradiant decay from $\nDown$ to $\nUp$ that increase this quantity \cite{hu2017PRA,swan2018PRL}. For smaller $R$, the larger value of $\delta_{\nDown}$ strongly suppresses dissipation relative to dispersive interactions (Eq.~(\ref{eqn:coup_strength_main})), leading to improved squeezing, i.e. smaller values of $\xi_\text{R}^2$. However, for fixed $\abs{\beta}^2$, the absolute squeezing rate $N\chi_{\nDown}$ also decreases with larger $\delta_{\nDown}$ (Eq.~(\ref{eqn:coup_strength_main})), leading to slower squeezing dynamics. Therefore, as summarized in the inset, smaller $R$ values enable greater metrological gain, but the time taken for squeezing also increases when $\abs{\beta}^2$ is fixed. 

The population dynamics at the different momentum centers reveal the effect of superradiance. Figure~\ref{fig:diffR_sameBeta}(b) shows the evolution of populations in $\nDown,\nUp$ for the case of $R=0.2$. The rapid decrease (increase) in $\nDown$ ($\nUp$) population reflects superradiant decay on the $\nDown\rightarrow\nUp$ transition. Further, the MCM enables an  investigation of the leakage to centers outside the spin manifold, highlighting the power of this technique. We denote the first $k$ centers higher than $\nUp$ as $n_{+k}$, and the first $k$ centers lower than $\nDown$ as $n_{-k}$.  The MCM reveals that a small number of atoms ($<10$) are lost to $n_{+1}$ during the squeezing dynamics, as seen in Fig.~\ref{fig:diffR_sameBeta}(c) for the various $R$ values. However, the excellent agreement between the TCM and MCM results in Fig.~\ref{fig:diffR_sameBeta}(a) indicates that in this parameter regime, the centers $\nDown,\nUp$ can be effectively treated as a closed two-level spin-$1/2$ manifold. 

\subsection{Squeezing faster and faster}

\begin{figure}[!htb]
    \centering
    \includegraphics[width=\columnwidth]{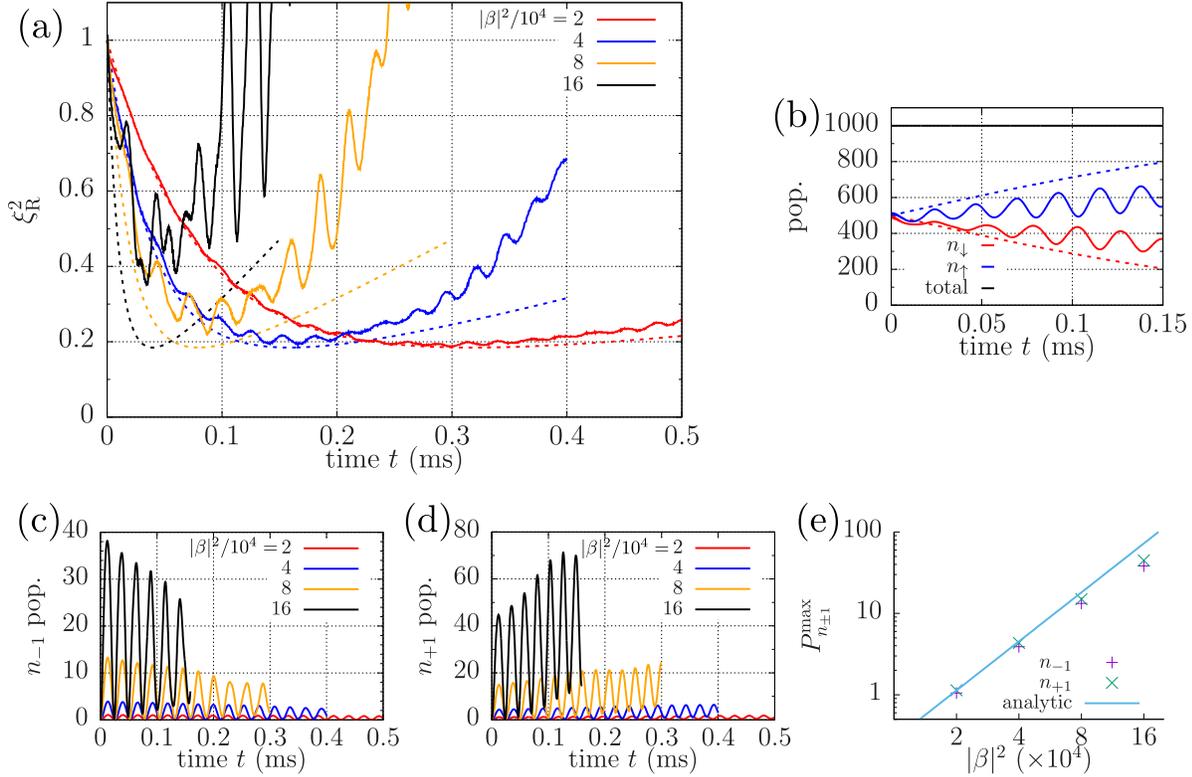}
    \caption{Squeezing faster and faster. (a) Evolution of $\xi_\text{R}^2$ for $\abs{\beta}^2/10^4 = 2,4,8,16$. (b) Population in $\nDown$, $\nUp$ for $\abs{\beta}^2/10^4=16$, with total population in all centers adding up to $N=10^3$. Solid (dashed) lines represent MCM (TCM) results. (c-d) Population in, respectively, $n_{-1}$ and $n_{+1}$ centers, for various drive strengths. (e) Comparison of simulated $n_{\pm 1}$ populations to analytic result of Rabi oscillation model (see Text). In this panel, $N=10^3$, $C=10$ and $R=0.05$. Six centers, $n_{-2}, n_{-1}, \nDown, \nUp, n_{+1}, n_{+2}$, were tracked in the MCM simulations with very low populations in $n_{\pm 2}$.   } 
    \label{fig:diffBeta_sameR}
\end{figure}

A simple two-level model, such as the TCM, would predict that the squeezing rate can be arbitrarily increased by simply pumping in more laser power so that $\abs{\beta}^2$ is increased. Figure~\ref{fig:diffBeta_sameR}(a) explores the evolution of $\xi_\text{R}^2$ in the case $C=10$, $R=0.05$ ($\delta_{\nDown}/2\pi=1$ MHz) for different values of $\abs{\beta}^2/10^4$ in the range $2-16$. As expected, the TCM (dashed) predicts that $\xi_\text{R}^2$ attains the same minimum value faster when $\abs{\beta}^2$ is increased. However, the MCM results (solid) present a different narrative: As $\abs{\beta}^2$ increases, $\xi_\text{R}^2$ indeed attains its minimum faster, but this value also increases, signaling a degradation of squeezing. In fact, the metrological gain $\xi_\text{R}^{-2}$ drops by $\sim 3$ dB (factor of 2) as $\abs{\beta}^2$ increases from $2 \times 10^4$ to $16 \times 10^4$.    

Large oscillations in the MCM curves as $\abs{\beta}^2$ is increased indicates the breakdown of the two-state model. A study of the population dynamics at the different centers confirms this breakdown. As seen in Fig.~\ref{fig:diffBeta_sameR}(b), although the populations in $\nDown$ ($\nUp$) follow the general decreasing (increasing) trend expected from $\nDown \rightarrow \nUp$ superradiant decay, the TCM and MCM population transients significantly differ in the case of strong driving ($\abs{\beta}^2/10^4=16$). Further, the MCM transients display pronounced oscillations with a frequency $\sim 8 \omega_r$, corresponding to the relative detuning between the $\nDown\leftrightarrow\nUp$ and $\nUp\leftrightarrow n_{+1}$, $n_{-1}\leftrightarrow \nDown$ transitions. 

Giant population oscillations in $n_{\pm1}$, shown in Fig.~\ref{fig:diffBeta_sameR}(c-d), confirm the significant participation of these centers in the dynamics as $\abs{\beta}^2$ increases.  A simple Rabi oscillation model qualitatively explains the occupation of these states: The coherent superposition of the $\nDown,\nUp$ centers serves as a large collective spin that sources mode 2. Both cavity modes, 1 and 2, are now macroscopically occupied and drive two-photon Rabi oscillations between $\nDown\leftrightarrow n_{-1}$ and $\nUp \leftrightarrow n_{+1}$ with approximate two-photon detuning $8\omega_r$. We find that the maximum population $P_{n_{\pm1}}^{\text{max}}$ in $n_{\pm 1}$ predicted by this model is given by (see \ref{sec:rabi_osc})
\begin{equation}
    P_{n_{\pm1}}^{\text{max}} \approx \frac{N}{2}\left(\frac{N\chi_{\nDown}}{8\omega_r}\right)^2.
    \label{eqn:pop_pm1}
\end{equation}
Figure~\ref{fig:diffBeta_sameR}(e) compares the first oscillation peak in the $n_{\pm1}$ populations with the analytic formula Eq.~(\ref{eqn:pop_pm1}). For small occupations (small $\abs{\beta}^2$), the formula agrees very well with the simulations, whereas the discrepancy becomes about a factor of 2 at the largest occupation ($\abs{\beta}^2=16 \times 10^4$). In this strong driving regime, the coherence that develops between $\nUp,n_{+1}$ and $\nDown,n_{-1}$ is no longer negligible and modifies the field in mode 2 considerably, leading to the breakdown of the simple Rabi oscillation picture presented here (\ref{sec:rabi_osc}). 

We also note that the amplitude of population oscillations in $n_{-1}$ ($n_{+1}$) decreases (increases) over time, as evident in Fig.~\ref{fig:diffBeta_sameR}(c) (Fig.~\ref{fig:diffBeta_sameR}(d)). The reason is that $\nDown\rightarrow\nUp$ superradiant decay irreversibly removes (introduces) atoms participating in the $\nDown\leftrightarrow n_{-1}$ ($\nUp\leftrightarrow n_{+1}$) Rabi cycles. Finally, the troughs, and consequently the peaks, of the oscillations in $n_{+1}$ population are clearly seen to rise with time because of the direct cavity mediated decay on the $\nUp\rightarrow n_{+1}$ transitions superposed on the Rabi oscillations. Similarly, the interplay of Rabi oscillations and cavity mediated decay on the $n_{-1}\rightarrow \nDown$ transition causes the troughs in the $n_{-1}$ oscillations to deviate from zero over time.

\subsection{Effect of momentum width}

\begin{figure}[!htb]
    \centering
    \includegraphics[width=\columnwidth]{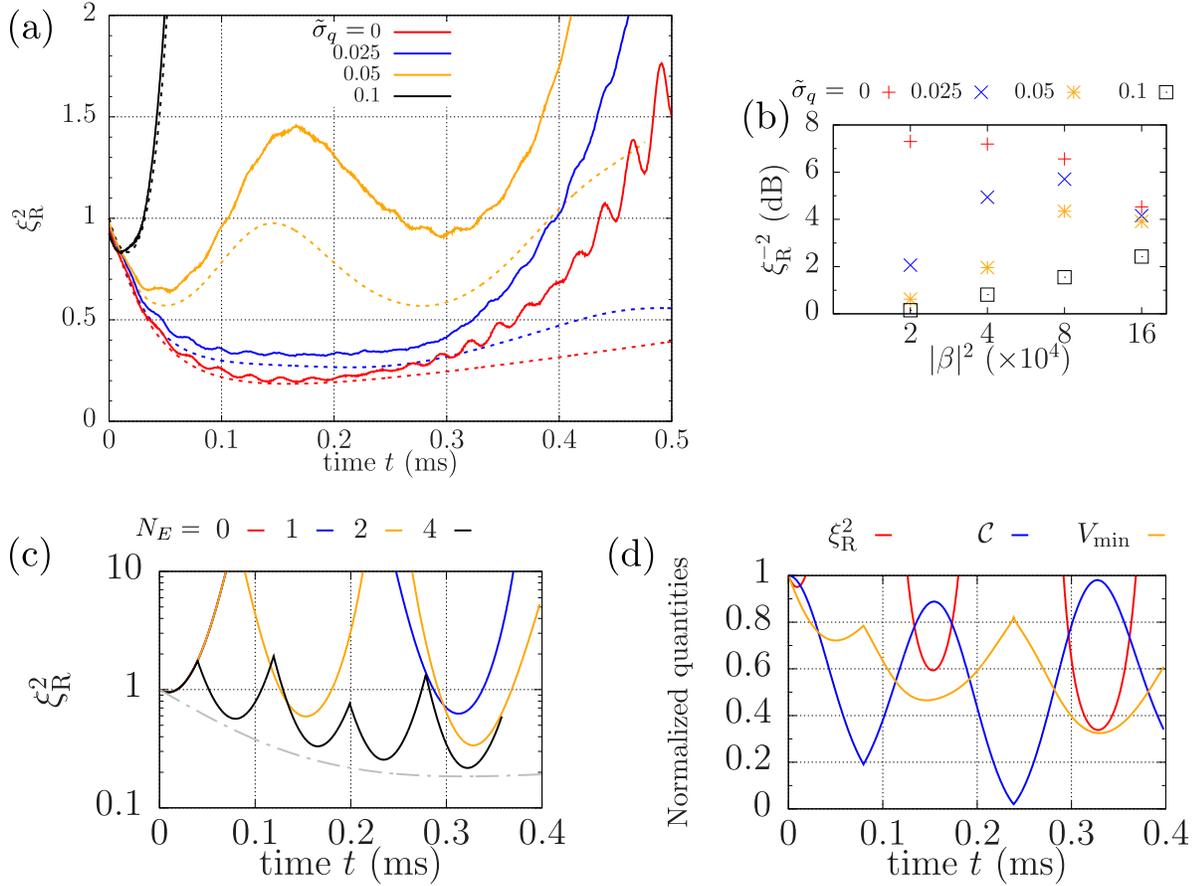}
    \caption{Squeezing in the presence of momentum width. (a) Evolution of $\xi_\text{R}^2$ in the case of $\abs{\beta}^2/10^4=4$ for $\td{\sigma}_q=0,0.025,0.05,0.1$. Solid (dashed) lines represent MCM (TCM) results. (b) Maximum metrological gain as a function of drive strength for different $\td{\sigma}_q$ values. (c) Evolution of $\xi_\text{R}^2$ in the TCM for $\td{\sigma}_q=0.1$ and $\abs{\beta}^2/10^4 = 2$ when $N_\text{E}=0,1,2,4$ echo pulses are inserted. The gray broken line shows the $\td{\sigma}_q=0$ case with no echoes. (d) Evolution of the constituents, $C$ and $V_\text{min}$ of $\xi_\text{R}^2$ in the TCM when $N_\text{E}=2$ echo pulses are inserted. Other details are the same as in Fig.~\ref{fig:diffBeta_sameR}. } 
    \label{fig:diffSig}
\end{figure}

We now consider the case when the atomic cloud has non-zero momentum width. For this study, we use the parameters from Fig.~\ref{fig:diffBeta_sameR}, i.e. $C=10$ and $\abs{\beta}^2/10^4 = 2,4,8,16$. Fig.~\ref{fig:diffSig}(a) shows the evolution of $\xi_{\text{R}}^2$ for $\td{\sigma}_q = 0, 0.025, 0.05$ and $0.1$ in the case when $\abs{\beta}^2/10^4=4$. In this panel, the solid and dashed curves respectively indicate the MCM and TCM models. Three trends can be observed from this figure: (T1) When the rate of squeezing is fast relative to the dephasing ($\propto \td{\sigma}_q$), the $\xi_\text{R}^2$ transient is similar (blue) to the zero width case (red) while the minimum value attained is greater indicating slight degradation of squeezing. (T2) For larger momentum width, the $\xi_\text{R}^2$ transient displays oscillatory behavior signifying competition between squeezing and dephasing (orange). (T3) As the width increases further and dephasing dominates, $\xi_\text{R}^2$ initially decreases slightly but then steeply increases to values well above unity, signaling rapid loss of squeezing (black). 

These trends are summarized in Fig.~\ref{fig:diffSig}(b), where the maximum metrological gain achievable is plotted as a function of $\abs{\beta}^2$ for different values of $\td{\sigma}_q$. The $\td{\sigma}_q=0$ case (red) reflects the study performed in  Fig.~\ref{fig:diffBeta_sameR} and shows that very strong driving lead to loss of squeezing as a result of coupling to other momentum centers. At the other extreme is the case of $\td{\sigma}_q=0.1$ (black), where rapid dephasing leads to a complete loss of squeezing for weak driving, and barely observable squeezing ($\sim 2$ dB) even for very strong driving. For intermediate widths $\td{\sigma}_q=0.025,0.05$ (blue, orange), the squeezing suffers at both ends, with dephasing restricting the squeezing at weak driving, and coupling to other centers serving as a limitation at very strong driving. For these widths, an optimum drive strength therefore exists where the metrological gain is maximized, as reflected by the variation of the gain for the four cases of $\abs{\beta}^2$ considered here.   

As Fig.~\ref{fig:diffSig}(a) exemplifies, we observe that the TCM (dashed) typically qualitatively reproduces the features seen in the MCM (solid) when studying the effect of momentum width. Except at very strong driving, the TCM and MCM agree reasonably well in the (T1) cases until the minimum squeezing time, after which the MCM rises very steeply compared to the TCM. In the (T2) cases, both models capture the oscillatory behavior but can be very different quantitatively. Finally, both models agree very well in the (T3) case. The difference in the two models is not only because of the extra momentum centers tracked by the MCM, but also because of the approximations used in solving for the dynamics in these models. In the TCM model, we force all non-trivial three-atom correlations to zero using a systematic truncation scheme (\ref{sec:eom_cumulant}). However, the MCM is a TWA-style approach that can, in general, capture the build-up of non-trivial three-atom correlations, which should be anticipated in an interacting system such as the one considered here. As an example, the general steep increase of the MCM curves after the minimum squeezing time in the (T1) cases is a manifestation of the effect of three-atom correlations, also visible in the cases plotted in Fig.~\ref{fig:diffBeta_sameR}(a). On the other hand, the superposed oscillations at frequency $\sim 8\omega_r$ are a result of coupling to the $n_{\pm 1}$ momentum centers.

The dephasing-induced degradation of squeezing can in fact be reversed. To elucidate this point, we consider the case of $\abs{\beta}^2/10^4=2$ and $\td{\sigma}_q=0.1$, a situation where achieving squeezing is seemingly hopeless because of weak driving and rapid dephasing (red curve in Fig.~\ref{fig:diffSig}(c)). As a minimal toy model to illustrate our protocol, we consider the TCM and interrupt the squeezing dynamics with a series of `instantaneous' echo pulses (\ref{sec:tcm_rot}). In a frame rotating at $4\omega_r$, the axis of rotation for these echoes is the same as that of the initial $\pi/2$-pulse used for preparing the equal superposition of the $\nDown=0,\nUp=1$ centers. Figure~\ref{fig:diffSig}(c) shows the evolution of $\xi_\text{R}^2$ when $N_\text{E}=0,1,2,4$ echo pulses are inserted during the course of the squeezing dynamics. The gray broken line shows the evolution of $\xi_\text{R}^2$ when $\td{\sigma}_q=0$. The timing of the $N_\text{E}>0$ echo pulses are such that they approximately divide the time to achieve the minimum $\xi_\text{R}^2$ in the $\td{\sigma}_q=0$ case ($\sim 0.3$ ms) into a sequence of $T,2T,\ldots,2T,T$ segments, where the number of $2T$ segments is $N_\text{E}-1$.  The insertion of echo pulses leads to a revival of $\xi_\text{R}^2$ as it periodically attains minima $<1$ as the spins re-phase after an echo pulse is applied. Increasing the number of such echoes prevents $\xi_\text{R}^2$ from blowing up to very large values at any point during its evolution and also maintains the periodically attained minima close to the $\td{\sigma}_q=0$ transient. 

The applicability of such a protocol to revive the squeezing parameter goes beyond only momentum pseudospins, and is useful on a variety of platforms where squeezing is desired in the presence of unavoidable on-site disorder, for example, in the case of NV centers. For a practical implementation using momentum pseudospins, the non-zero echo pulse duration ( $\gtrsim 2\pi/4\omega_r$ to avoid leakage to centers outside $\nDown=0,\nUp=1$) and the effect of momentum width on pulse efficiency \cite{Szigeti2012} have to be considered. Nevertheless, with suitable choice of parameters, we anticipate partial revivals in $\xi_\text{R}^2$ to be observable despite these deviations from our toy model.  

Finally, we investigate the constituent observables of the spin squeezing parameter to better understand this strong revival phenomenon. From Eq.~(\ref{eqn:wineland}), $\xi_\text{R}^2$ comprises of two observables, namely, $\mathcal{C}$ (Eq.~(\ref{eqn:ctrst})) and $V_\text{min}$ (Eq.~(\ref{eqn:vmin})). Figure~\ref{fig:diffSig}(d) plots the evolution of these observables as well as $\xi_\text{R}^2$ for the case of $N_\text{E}=2$. The re-phasing of the spins after each echo leads to the expected increase of $\mathcal{C}$. However, Fig.~\ref{fig:diffSig}(d) shows that this increase alone is not responsible for the strong revival of $\xi_\text{R}^2$. As the spins re-phase, $V_\text{min}$ also reaches its minima close to the times when $\mathcal{C}$ peaks, thereby leading to sharp dips in $\xi_\text{R}^2$.

\subsection{\label{sec:gap_protect}Collective physics with a many-body energy gap}

\begin{figure}[!htb]
    \centering
    \includegraphics[width=\columnwidth]{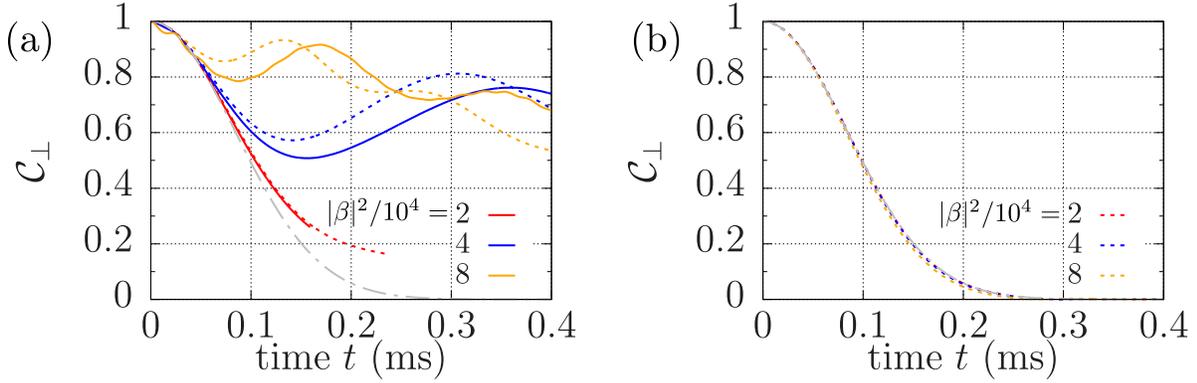}
    \caption{Manifestation of a many-body energy gap. (a) Evolution of $\mathcal{C}_\perp$ for $\td{\sigma}_q=0.05$ for different values of $\abs{\beta}^2/10^4=2,4,8$. (b) TCM results using the same parameters as in (a), but with the gap Hamiltonian $\hat{H}_\text{G}$ turned off. The gray broken line in each case shows the decay of $\mathcal{C}_\perp$ under free evolution. Solid (dashed) lines represent MCM (TCM) results. Other details are the same as in Fig.~\ref{fig:diffBeta_sameR}.} 
    \label{fig:gap_protection}
\end{figure}

Apart from squeezing, yet another type of collective behavior manifests as a result of the cavity mediated atom-atom interactions. We consider the observable $\mathcal{C}_\perp$, defined as the normalized length of the projection of the Bloch vector on to the equatorial plane of the Bloch sphere. Mathematically, 

\begin{equation}
    \mathcal{C}_\perp = \frac{\sqrt{\ev{\hat{\mJ}_{\nDown}^x}^2 + \ev{\hat{\mJ}_{\nDown}^y}^2}}{N/2}.
    \label{eqn:cperp}
\end{equation}
Figure~\ref{fig:gap_protection}(a) plots the evolution of $\mathcal{C}_\perp$ in the case $\td{\sigma}_q=0.05$ for different values of $\abs{\beta}^2/10^4 = 2,4,8$. The TCM (dashed) and the MCM (solid) are in qualitative agreement in all cases and in quantitative agreement when dephasing dominates, i.e. for weak driving (red). The gray broken line shows the corresponding decay of $\mathcal{C}_\perp$ for freely evolving atoms, i.e. with no interactions, which obeys the analytical expression $\mathcal{C}_\perp (t)=e^{-\mu_d^2 t^2}$, where $\mu_d = 4\sqrt{2}\omega_r \td{\sigma}_q$. Clearly, interactions lead to an observably slow decay of contrast compared to the free evolution case.

The effective Hamiltonian, Eq.~(\ref{eqn:h_eff_2_truncated}), provides insight into the slow decay of $\mathcal{C}_\perp$ in the presence of interactions. We note that for any $n$,
\begin{equation}
    \hat{\mJ}_n^- \hat{\mJ}_n^+ = \mbf{\hat{J}_n}\cdot\mbf{\hat{J}_n} - \hat{\mJ}_n^z \hat{\mJ}_n^z - \hat{\mJ}_n^z. 
\end{equation}
We introduce the many-body gap Hamiltonian, $\hat{H}_\text{G} = \hbar\chi_{\nDown} \mbf{\hat{J}_{\nDown}}\cdot\mbf{\hat{J}_{\nDown}}$.  The initial uncorrelated many-body state can be visualized as a coherent spin state in the equatorial plane of the Bloch sphere corresponding to the maximum quantum number $J_{\nDown} = N/2$ associated with the operator  $\mbf{\hat{J}_{\nDown}}\cdot\mbf{\hat{J}_{\nDown}}$.  In other words, this initial state satisfies
\begin{equation}
\ev{\mbf{\hat{J}_{\nDown}}\cdot\mbf{\hat{J}_{\nDown}}(0)} = \frac{N}{2}\left(\frac{N}{2}+1\right), \; 
\mathcal{C}_\perp(0) = 1.
\end{equation}
The first term of $\hat{H}_\text{eff}$ in Eq.~(\ref{eqn:h_eff_2_truncated}) is not collective, causing dephasing of individual spins that leads to shortening of the mean spin length and populates shells of lower $J_{\nDown}$. The presence of $\hat{H}_\text{G}$ introduces an energy penalty for populating shells of lower $J_{\nDown}$. Specifically, $\hat{H}_\text{G}$ dictates that
\begin{equation}
    \hat{H}_\text{G} \ket{J_{\nDown},M_{J_{\nDown}}} = \hbar\chi_{\nDown} J_{\nDown}\left(J_{\nDown}+1\right),
\end{equation}
implying that the transition to a lower shell, $J_{\nDown}\rightarrow J_{\nDown}-1$, incurs an energy penalty  
\begin{equation}
    \abs{\Delta E_{\left(J_{\nDown}\rightarrow J_{\nDown}-1\right)}} = 2 \hbar\chi_{\nDown} J_{\nDown}.
\end{equation}
As a result, individual atom dephasing is slowed down, leading to slower decay of $\mathcal{C}_\perp$. 

We verify this qualitative explanation in
Fig.~\ref{fig:gap_protection}(b), where we study the dynamics of
$\mathcal{C}_\perp$ under the TCM with the gap Hamiltonian
$\hat{H}_\text{G}$ turned off. The decay of $\mathcal{C}_\perp$ is
then in excellent agreement with the free evolution case, although
interactions are present through the remaining terms in
Eq.~(\ref{eqn:h_eff_2_truncated}) and the dissipative term of
Eq.~(\ref{eqn:me_truncated}).

Investigations with the TCM indicate that the presence of the gap Hamiltonian 
$\hat{H}_\text{G}$ is an advantage from a metrology perspective. The slow decay of contrast 
leads to a smaller value for the minimum squeezing parameter $\xi_\text{R}^2$
compared to the case when $\hat{H}_\text{G}$ is turned off. Further,
the subsequent rise of $\xi_\text{R}^2$ after the minimum value is
attained is slowed down when $\hat{H}_\text{G}$ is
present. We note that the non-zero momentum spread is an intrinsic
source of dephasing in a Bragg interferometer, and the cavity-mediated
interactions we engineer naturally provide a many-body gap protection
that suppresses this dephasing.

In general, our results are consistent with other examples that
confirm that the presence of a many-body gap arising from correlations
can supresses adverse effects of single-atom
decoherence~\cite{norcia2018Sci} and potentially contribute to extending the
coherence time for precision metrology. This ability to engineer
many-body correlations driven either by mediated interactions or
particle statistics represents an emerging paradigm for advanced
metrology~\cite{lucchesi2019arXiv}.

\section{\label{sec:conc}Conclusion}

We have proposed and analyzed in detail a scheme for squeezing directly on momentum pseudospins using cavity-mediated atom-atom interactions. Implementing our scheme does not require any experimental overhead beyond what is necessary to operate Bragg intereferometers in a cavity. Since our scheme relies on emission and absorption of a cavity photon, it is only applicable to states separated by $2\hbar k$. Nevertheless, the squeezing can be transferred to higher diffraction orders by subsequently applying large momentum transfer pulses \cite{chiow2011PRL,swing2018PRL}. For studying various aspects of the problem, we have focused on the ${}^{1}\text{S}_0 - {}^{3}\text{P}_1$ transition in ${}^{88}\text{Sr}$ as an example, working in parameter regimes where $< 10$ dB of metrological gain is achievable in a few hundred microseconds to a few milliseconds based on the driving strength. While more than sufficient for a proof-of-principle experiment, we expect that with suitable choice of parameters- small momentum width, small ratios of dissipative to dispersive interactions ($R = \kappa/2\delta_{\nDown}$) and moderately strong driving strengths, $\gtrsim 10$ dB of metrological gain can be achieved. Such parameters are within the reach of current technology: Velocity selection techniques are able to provide clouds with $\td{\sigma}_q \leq 0.01$. The $R$ value can be tuned to smaller values by detuning the drive laser farther away from the cavity resonance. Strong driving at large detunings is not a problem since modern lasers are able to deliver orders of magnitude more power than the hundreds of microwatts required in our proof-of-principle parameter regimes. In addition to squeezing, the same experimental setup can also be used to demonstrate and explore collective physics associated with the opening of a many-body energy gap by measuring a different observable, namely the contrast $\mathcal{C}_\perp$ (Eq.~(\ref{eqn:cperp})).

In addition to superradiant decay, single atom free-space scattering (FSS) also degrades the squeezing. Superradiance, being collectively enhanced, is the dominant source of degradation in most of the parameter regimes we have considered (\ref{sec:fss}) and therefore we have only focused on this dissipation mechanism. The parameter regime where superradiance dominates FSS is $R^2 \gg 1/NC$ (\ref{sec:fss}), and therefore, FSS is not important when large atom numbers are used such that this inequality is satisfied. Nevertheless, FSS can be straightforwardly included in both the simulation models demonstrated here with very little computational overhead by accounting for the corresponding Lindblad terms. The scaling of the multi-center model remains linear in atom number since FSS occurs independently for each atom.  

While in principle the $R$ value can be made arbitrarily small to suppress superradiance and greatly improve the squeezing, with fixed atom number the power required to squeeze at a specified rate $Q$ rapidly increases as $1/R^3$ (Eq.~(\ref{eqn:q_power_relation})), motivating considerations of elegant related schemes that are not as sensitive to superradiance. Recent schemes developed for squeezing on optical clock transitions circumvent this problem by either squeezing faster using a twist and turn scheme achieved by introducing a resonant drive \cite{hu2017PRA} or by an unconventional choice of initial state that drives the squeezing in a spin component orthogonal to that affected by superradiant decay \cite{swan2018PRL}. The former can be implemented on momentum pseudospins using an additional pair of resonant Bragg lasers injected, for example, one free spectral range away from the cavity mode used for squeezing. The latter scheme requires an initial state with two ensembles pointing along opposite directions in the equatorial plane of the Bloch sphere. It can be implemented by launching two clouds with equal atoms which are initially in the $\nDown$ and $\nUp$ states respectively and applying a common $\pi/2$-pulse to rotate them to the equatorial plane. However, in either case, a careful study of the effects of momentum width and potential leakage to other momentum centers has to be performed. The techniques developed in this paper can be readily used to undertake such a study. The latter scheme, combined with differential rotations on the two ensembles \cite{aguila2018NJP}, can potentially be used to implement an entangled atom Bragg gradiometer. 

Finally, several mature atomic and atom-like platforms are beginning to demonstrate exotic many-body phenomena such as discrete time crystals \cite{zhang2017Nat,choi2017Nat}, many-body localization \cite{kaufman2016Sci,lukin2019Sci} and dynamical phase transitions \cite{jurcevic2017PRL,smale2018arxiv}. Bragg interferometers operating in cavities open avenues for engineering interactions, and the theoretical techniques we have developed in this paper can be used to explore the complex interplay of interactions, losses, disorder and global state rotations in other configurations involving momentum pseudospins. 

\ack{We would like to thank Baochen Wu, James Thompson, Luca
  Pezz{\`e}, Augusto Smerzi, John Cooper and Robert Lewis-Swan for
  useful discussions and constructive comments on the
  manuscript. A.~S. and M.~H. acknowledge financial support from NSF
  PFC Grant No. PHY 1734006 and DARPA Extreme Sensing. L.~S.,
  M.~L.~C. and N.~P. acknowledge financial support from INFN and the
  Italian Ministry of Education, University and Research (MIUR) under
  the Progetto Premiale ``Interferometro Atomic'' and PRIN
  2015. N.~P. acknowledges financial support from European Research
  Council, Grant No. 772126 (TICTOCGRAV). M. L. C. is grateful to JILA
  for the warm welcome and conducive research environment during the
  Visiting Fellowship, when part of this work was carried out.}

\appendix

\section{Derivation of effective master equation for atom-atom interactions} 

\subsection{\label{sec:ad_exc}Adiabatic elimination of the excited state}

We first transform to an interaction picture rotating at the drive frequency $\omega_l$ with free evolution Hamiltonian $H_f^\text{(1)} = \sum_j \hbar \omega_l/2(\ket{e}_j\bra{e}-\ket{g}_j\bra{g}) + \sum_{s} \omega_l \hat{a}_s^\dag \hat{a}_s  $. The resulting interaction picture Hamiltonian is  
\begin{eqnarray}
\hat{H}_I^\text{(1)} &=& \sum_{j=1}^N \left( \frac{\hat{p}_j^2}{2M} + \frac{\hbar \Delta_l}{2}(\ket{e}_j\bra{e}-\ket{g}_j\bra{g}) \right) - \sum_{s=1}^2 \hbar \Delta_{cl} \hat{a}_s^\dag \hat{a}_s \nonumber\\
&+& \sum_{j=1}^N \sum_{s=1}^2 \frac{\hbar g}{2}\left( \hat{a}_s e^{i k_s \hat{z}_j}\ket{e}_j\bra{g} + \hat{a}_s^\dag e^{-i k_s \hat{z}_j}\ket{g}_j\bra{e}  \right) \nonumber\\
&+& \hbar \sqrt{\kappa} \left( \alpha  \hat{a}_1^\dag + \alpha^* \hat{a}_1\right).
\label{eqn:interaction_ham_1}
\end{eqnarray}
The coherence operator $\ket{e}_j\bra{g}$ satisfies the equation 
\begin{equation}
    \frac{d}{dt}\ket{e}_j\bra{g} = i\Delta_l \ket{e}_j\bra{g} - i\frac{g}{2}\sum_{s=1}^2\hat{a}_s^\dag e^{-i k_s \hat{z}_j} \left( \ket{e}_j\bra{e}-\ket{g}_j\bra{g} \right).
\end{equation}
In a far-detuned regime, we can set $\ket{e}_j\bra{e}-\ket{g}_j\bra{g} \approx -1$. We then transform to the cavity frame by substituting $\hat{a}_s^\dag = \hat{a}_s^{\dag,\text{(c)}} e^{i\Delta_{cl}t}$, $\ket{e}_j\bra{g} = \ket{e}_j\bra{g}^\text{(c)} e^{i\Delta_{cl}t}$  and adiabatically eliminate $\ket{e}_j\bra{g}^\text{(c)}$ to get 
\begin{equation}
    \ket{e}_j\bra{g}^\text{(c)} \approx -\frac{g}{2\Delta_c}\sum_{s=1}^2 \hat{a}_s^{\dag,\text{(c)}} e^{-i k_s \hat{z}_j}.
    \label{eqn:coherence_elim}
\end{equation}
In the drive frame, the annihilation operator for a mode $s$ satisfies the equation 
\begin{equation}
    \frac{d}{dt}\hat{a}_s = -\left( \frac{\kappa}{2}-i\Delta_{cl}\right)\hat{a}_s -i\frac{g}{2}\sum_{j=1}^N e^{-i k_s \hat{z}_j} \ket{g}_j\bra{e} -i\sqrt{\kappa} \alpha \delta_{s,1} + \hat{F}_s,
\end{equation}
where $\hat{F}_s$ is the noise operator associated with coupling to the modes outside the cavity. Using the hermitian conjugate of the expression, Eq.~(\ref{eqn:coherence_elim}), leads to 
\begin{eqnarray}
    \frac{d}{dt}\hat{a}_s &\approx& -\left( \frac{\kappa}{2}-i\Delta_{cl}\right)\hat{a}_s +i\frac{g^2}{4\Delta_c}\sum_{j=1}^N \sum_{s^\prime=1}^{2} \hat{a}_{s^\prime} e^{-i (k_s-k_{s^\prime}) \hat{z}_j} \nonumber\\
    &-&i\sqrt{\kappa} \alpha \delta_{s,1} + \hat{F}_s.
\end{eqnarray}
These equations can be obtained from the effective Hamiltonian 
\begin{eqnarray}
\hat{H}_I^\text{(2)} &=& \sum_{j=1}^N \frac{\hat{p}_j^2}{2M}  - \sum_{s=1}^2 \hbar \left(\Delta_{cl} + \frac{N g^2}{4\Delta_c} \right) \hat{a}_s^\dag \hat{a}_s \nonumber\\
&-& \sum_{j=1}^N \frac{\hbar g^2}{4\Delta_c}\left( \hat{a}_1^\dag \hat{a}_2 e^{-i k_\text{eff} \hat{z}_j} + \hat{a}_2^\dag \hat{a}_1 e^{i k_\text{eff} \hat{z}_j}\right) \nonumber\\
&+& \hbar \sqrt{\kappa} \left( \alpha  \hat{a}_1^\dag + \alpha^* \hat{a}_1\right).
\label{eqn:interaction_ham_2}
\end{eqnarray}
Here $k_\text{eff}=k_1-k_2=2k$ is the effective wavevector. The cavity resonance is now shifted by $-Ng^2/4\Delta_c$ because of the presence of the atoms. Modifying the drive frequency $\omega_l \rightarrow \omega_l - Ng^2/4\Delta_c$ returns the detuning to $\Delta_{cl}$.

\subsection{\label{sec:mode2_elim}Elimination of the cavity field $\hat{a}_2$}

We follow a similar procedure to that presented in Appendix~C of Ref. \cite{shankar2017PRA}. We split the master equation, Eq.~(\ref{eqn:master_eqn_dtwa}), into system, reservoir as well as system-reservoir Liouvillians. These terms are given by 
\begin{eqnarray}
    \lv_S \rho_{a-c} &=& -i\left[ \sum_{j=1}^N \frac{\hat{p}_j^2}{2M\hbar}, \rho_{a-c} \right], \nonumber\\
    \lv_R \rho_{a-c} &=& -i\left[ -\Delta_{cl} \hat{a}_2^\dag \hat{a}_2, \rho_{a-c} \right] + \kappa \mD[\hat{a}_2]\rho_{a-c}, \nonumber\\
    \lv_{SR} \rho_{a-c} &=& = -i\left[-\sum_{j=1}^N \frac{g^2}{4\Delta_c}\left( \beta^* \hat{a}_2 e^{-i k_\text{eff} \hat{z}_j} + \beta \hat{a}_2^\dag e^{i k_\text{eff} \hat{z}_j}\right),\rho_{a-c}\right].\nonumber\\
\end{eqnarray}
We first transform to an interaction picture with $\lv_0 = \lv_S + \lv_R$. We then have
\begin{equation}
    \dot{\td{\rho}}_{a-c} = \td{\lv}_{SR}\td{\rho}_{a-c},
    \label{eqn:int_pic_me}
\end{equation}
where $\td{\rho}_{a-c} = e^{-\lv_0 t}\rho_{a-c}$ and $\td{\lv}_{SR} = e^{-\lv_0 t}\lv_{SR}e^{\lv_0 t}$. We integrate Eq.~(\ref{eqn:int_pic_me}) and substitute the formal solution for $\td{\rho}_{a-c}(t)$ in the same equation to get 
\begin{equation}
    \dot{\td{\rho}}_{a-c} = \td{\lv}_{SR}\td{\rho}_{a-c}(0) + \int_0^t dt' \td{\lv}_{SR}(t)\td{\lv}_{SR}(t')\td{\rho}_{a-c}(t').
\end{equation}
We assume that mode 2 acts as a reservoir in the vacuum state, i.e. the reservoir density matrix $R_0 = \ket{0}\bra{0}$. At $t=0$, the initial uncorrelated state is $\td{\rho}_{a-c}(0) = \td{\rho}_a(0)R_0$, where $\td{\rho}_a(0)$ is the density matrix for the atomic ensemble. We then use a decorrelation approximation to write $\td{\rho}_{a-c}(t)\approx \td{\rho}_a(t)R_0$ for later times, and trace out mode 2 as 
\begin{eqnarray}
    \dot{\td{\rho}}_a = \Tr_R\left[ \td{L}_{SR}(t)\td{\rho}_a(0)R_0\right] 
    + \int_0^t dt' \Tr_R[\td{\lv}_{SR}(t)\td{\lv}_{SR}(t')\td{\rho}_a(t')R_0].
    \label{eqn:me_start}
\end{eqnarray}
The first term vanishes because $\ev{\hat{a}_2} = \ev{\hat{a}_2^\dag} = 0$ in the vacuum state.

Next, we find the time evolution equations governing the superoperators associated with mode 2 that enter $\td{\lv}_{SR}$, namely  $\td{\hat{a}}_2 \otimes \hat{I}$, $\td{\hat{a}}_2^\dag \otimes \hat{I}$, $\hat{I}\otimes(\td{\hat{a}}_2)^T$ and $\hat{I}\otimes(\td{\hat{a}}_2^\dag)^T$. Here $\hat{I}$ is the identity operator, i.e. $\hat{I}\ket{n}=\ket{n}$ for any Fock basis vector $\ket{n}$. The notation $\hat{A}\otimes(\hat{B})^T$ is to be understood as the operation $\hat{A}\ket{n}\bra{m}\hat{B}$ for a vector $\ket{n}\bra{m}$ in the Liouville space of mode 2 \cite{shankar2017PRA}. These equations are found to be
\begin{eqnarray}
    \frac{d}{dt}\td{\hat{a}}_2\otimes \hat{I} &=& -\left(\frac{\kappa}{2}-i\Delta_{cl} \right) \td{\hat{a}}_2\otimes \hat{I} \nonumber\\
    \frac{d}{dt}\hat{I}\otimes(\td{\hat{a}}_2)^T &=& \left(\frac{\kappa}{2}+i\Delta_{cl} \right)\hat{I}\otimes(\td{\hat{a}}_2)^T -\kappa \left( \td{\hat{a}}_2\otimes \hat{I}\right). 
\end{eqnarray}
The solution to this coupled set of differential equations is 
\begin{eqnarray}
    \td{\hat{a}}_2\otimes \hat{I}(t) &=& \left(\hat{a}_2\otimes \hat{I}\right) e^{-\left(\frac{\kappa}{2}-i\Delta_{cl}\right)t} \nonumber\\
    \hat{I}\otimes(\td{\hat{a}}_2)^T(t) &=& \left[ \hat{I}\otimes(\hat{a}_2)^T - \hat{a}_2\otimes \hat{I}\right] e^{\left(\frac{\kappa}{2}+i\Delta_{cl}\right)t}
    + \left(\hat{a}_2\otimes \hat{I}\right) e^{-\left(\frac{\kappa}{2}-i\Delta_{cl}\right)t}. \nonumber\\
    \label{eqn:superop_time_evolve}
\end{eqnarray}
Hermitian conjugation of these two equations yields the expressions for $\td{\hat{a}}_2^\dag \otimes \hat{I} (t)$ and $\hat{I}\otimes(\td{\hat{a}}_2^\dag)^T (t)$.

For brevity, we denote $\hat{S}_j \equiv e^{i\kE \hat{z}_j}$. From Eqs.~(\ref{eqn:me_start}) and (\ref{eqn:superop_time_evolve}), we arrive at 
\begin{eqnarray}
    \dot{\td{\rho}}_a = -\left(\frac{g^2\abs{\beta}}{4\Delta_c}\right)^2 \sum_{j,j'=1}^N \int_0^t dt' 
    &&\left[\td{\hat{S}}_j^\dag(t) \td{\hat{S}}_{j'}(t')\td{\rho}_a(t') e^{-\left(\kappa/2-i\Delta_{cl}\right)(t-t')}\right.\nonumber\\
    &-& \td{\hat{S}}_j(t) \td{\rho}_a(t') \td{\hat{S}}_{j'}^\dag(t') e^{-\left(\kappa/2+i\Delta_{cl}\right)(t-t')}\nonumber\\
    &-& \td{\hat{S}}_{j'}(t') \td{\rho}_a(t') \td{\hat{S}}_j^\dag(t) e^{-\left(\kappa/2-i\Delta_{cl}\right)(t-t')}\nonumber\\
    &+& \left.\td{\rho}_a(t') \td{\hat{S}}_{j'}^\dag(t') \td{\hat{S}}_j(t) e^{-\left(\kappa/2+i\Delta_{cl}\right)(t-t')}\right].
    \label{eqn:me_born}
\end{eqnarray}

In arriving at Eq.~(\ref{eqn:me_born}), we have used the fact that the reservoir is approximately in the vacuum state to set $\ev{\hat{a}_2\hat{a}_2}=\ev{\hat{a}_2^\dag\hat{a}_2} = 0$ and $\ev{\hat{a}_2 \hat{a}_2^\dag} = 1$. The time evolution of the system operator $\td{\hat{S}}_j(t)$ is given by 
\begin{equation}
    \td{\hat{S}}_j(t) = \exp\left( i\frac{\hat{p}_j^2}{2M\hbar}t\right) e^{i\kE \hat{z}_j} \exp\left( -i\frac{\hat{p}_j^2}{2M\hbar}t\right).
\end{equation}

Once again, we introduce generalized population and coherence operators, but with an extra label $q$, as
\begin{equation}
    \hSig_{nm}^{j,q} = \ket{n,q}_j\bra{m,q},
\end{equation}
and expand the momentum shift operator $e^{i \kE \hat{z}_j}$ as 
\begin{equation}
 e^{i \kE \hat{z}_j} = \sum_{n=\infty}^{\infty}\int_{-\infty}^{\infty} dq \hSig_{n+1,n}^{j,q}. 
 \label{eqn:mom_shift_decomp}
\end{equation}
We then have 
\begin{equation}
    \td{\hat{S}}_j(t) = \sum_{n=\infty}^{\infty}\int_{-\infty}^{\infty} dq
    e^{i \Delta\omega_n(q) t} \hSig_{n+1,n}^{j,q},
\end{equation}
where we have introduced $\Delta\omega_n(q)=4\omega_r\left(1+2n+2\tdq\td{\sigma}_q\right)$. 
As an example, we explicitly write down the first term in Eq.~(\ref{eqn:me_born}):
\begin{eqnarray}
     -\left(\frac{g^2\abs{\beta}}{4\Delta_c}\right)^2 \sum_{j,j'}&&\int_0^t dt' 
     \sum_{n,n'} \int_{-\infty}^{\infty}dq \int_{-\infty}^{\infty}dq'
     \hSig_{n,n+1}^{j,q} \hSig_{n'+1,n'}^{j',q'} \td{\rho}_a(t')\times\nonumber\\
     &&\exp\left[-\left(\frac{\kappa}{2}-i\delta_n(q)\right)t\right]
     \exp\left[\left(\frac{\kappa}{2}-i\delta_{n'}(q')\right)t'\right],
     \label{eqn:bef_markov}
\end{eqnarray}
where $\delta_n(q) \equiv \Delta_{cl}-\Delta\omega_n(q)$.

The restriction $\td{\sigma}_q\ll1$ ensures that only operators associated with $q\ll\hbar\kE$ contribute to the dynamics. We further assume that for the momentum centers that significantly participate in the dynamics, the corresponding $\abs{\kappa/2-i\delta_n(q)}$ is sufficiently `large'. We will quantify this criterion self-consistently later on (see \ref{sec:markov_validity}). Then, the integral over $t'$ can be performed under a Markov approximation by setting $\td{\rho}_a(t') \approx \td{\rho}_a(t)$ to get 
\begin{eqnarray}
    -\sum_{j,j'} \sum_{n,n'} \int_{-\infty}^{\infty}dq \int_{-\infty}^{\infty}dq'
    \frac{\left(g^2\abs{\beta}/4\Delta_c\right)^2}{\frac{\kappa}{2}-i\delta_{n'}(q')}
    e^{-i\left(\Delta\omega_n(q)-\Delta\omega_{n'}(q')\right)t}
    \hSig_{n,n+1}^{j,q} \hSig_{n'+1,n'}^{j',q'} \td{\rho}_a(t). \nonumber\\
\end{eqnarray}
We repeat this calculation for the remaining three terms. We define the coherent and dissipative coupling strengths as 
\begin{eqnarray}
    \chi_n(q) = \left(\frac{g^2\abs{\beta}}{4\Delta_c}\right)^2 \frac{\delta_n(q)}{\kappa^2/4+(\delta_n(q))^2}, \;
    \Gamma_n(q) = \left(\frac{g^2\abs{\beta}}{4\Delta_c}\right)^2 \frac{\kappa/2}{\kappa^2/4+(\delta_n(q))^2},\nonumber\\
    \label{eqn:coup_strength}
\end{eqnarray}
and perform the reverse interaction picture transformation with $\lv_0 = -\lv_S$ to obtain an effective master equation governing the dynamics of $\rho_a$:
\begin{eqnarray}
    \dot{\rho}_a &=& \frac{1}{i\hbar} \left[ \sum_{j=1}^N  \sum_{n=-\infty}^{\infty}\int_{-\infty}^{\infty}dq\; \hbar\omega_n(q)\hSig_{nn}^{j,q},\;\rho_a\right] \nonumber\\
    &-&i\sum_{j,j'} \sum_{n,n'}\int_{-\infty}^{\infty}dq \int_{-\infty}^{\infty}dq' \chi_{n'}(q')
    \left( \hSig_{n,n+1}^{j,q}\hSig_{n'+1,n'}^{j',q'}\rho_a - \rho_a \hSig_{n',n'+1}^{j',q'}\hSig_{n+1,n}^{j,q} \right.\nonumber\\
    &&\hphantom{-i\sum_{j,j'} \sum_{n,n'}\int_{-\infty}^{\infty}dq \int_{-\infty}^{\infty}dq' \chi_n}
    \left. +\hSig_{n+1,n}^{j,q} \rho_a \hSig_{n',n'+1}^{j',q'} - \hSig_{n'+1,n'}^{j',q'} \rho_a \hSig_{n,n+1}^{j,q} \right)\nonumber\\
    &+&\sum_{j,j'} \sum_{n,n'}\int_{-\infty}^{\infty}dq \int_{-\infty}^{\infty}dq' \Gamma_{n'}(q')
    \left( \hSig_{n+1,n}^{j,q} \rho_a \hSig_{n',n'+1}^{j',q'} + \hSig_{n'+1,n'}^{j',q'} \rho_a \hSig_{n,n+1}^{j,q} \right.\nonumber\\
    &&\hphantom{+\sum_{j,j'} \sum_{n,n'}\int_{-\infty}^{\infty}dq \int_{-\infty}^{\infty}dq' \Gamma_n}
    \left. -\hSig_{n,n+1}^{j,q}\hSig_{n'+1,n'}^{j',q'}\rho_a - \rho_a \hSig_{n',n'+1}^{j',q'}\hSig_{n+1,n}^{j,q} \right).\nonumber\\
    \label{eqn:me_eff}
\end{eqnarray}

We make the simplifying assumption that $\chi_n(q)\approx\chi_n(0)\equiv\chi_n$, $\Gamma_n(q)\approx\Gamma_n(0)\equiv\Gamma_n$, that allows to pull $\chi_n,\Gamma_n$ outside the integrals. We find that this requirement constrains \begin{equation}
    \frac{\sigma_q}{\hbar\kE} \ll \underset{n}{\text{min}}\left( \frac{\delta_n}{16\omega_r}\right),
    \label{eqn:sigma_q_req2}
\end{equation}
where $\delta_n\equiv\delta_n(0)$ and the values of $n$ considered correspond to the centers that significantly participate in the dynamics. In deriving the simple expression in Eq.~(\ref{eqn:sigma_q_req2}), we have assumed that the dispersive interaction dominates, i.e. $\delta_n\gg\kappa/2$ for participating centers. For detunings $\delta_{\nDown} \gg 4\omega_r$, Eq.~(\ref{eqn:sigma_q_req1}) is clearly a more stringent requirement than Eq.~(\ref{eqn:sigma_q_req2}). 

Further, the simultaneous excitation and de-excitation of a pair of atoms is near-resonant only when the same centers are involved, i.e. $n=n'$. For $n=n'\pm 1$, the exchange process is energetically detuned by $8 \omega_r$\footnote{We note that ignoring terms with $n\neq n'$ amounts to assuming that rates of the order of $8\omega_r$ are `rapidly oscillating'. Therefore, the model we derive here is only valid for squeezing rates $N\chi_{\nDown} \ll 8\omega_r$, and cannot predict all features seen in the MCM in the strong driving regime (such as in Fig.~(\ref{fig:diffBeta_sameR}), also see \ref{sec:rabi_osc}) even when more than two centers are tracked.}. From these considerations, the effective master equation, Eq.~(\ref{eqn:me_eff}), can be written as 
\begin{eqnarray}
    \dot{\rho}_a &=& \frac{1}{i\hbar} \left[ \sum_j  \sum_n\int_{-\infty}^{\infty}dq\; \hbar\omega_n(q)\hSig_{nn}^{j,q},\;\rho_a\right] \nonumber\\
    &-&i\sum_{j,j'} \sum_n \chi_n \int_{-\infty}^{\infty}dq \int_{-\infty}^{\infty}dq' 
    \left( \hSig_{n,n+1}^{j,q}\hSig_{n+1,n}^{j',q'}\rho_a - \rho_a \hSig_{n,n+1}^{j',q'}\hSig_{n+1,n}^{j,q} \right.\nonumber\\
    &&\hphantom{-i\sum_{j,j'} \sum_n \int_{-\infty}^{\infty}dq \int_{-\infty}^{\infty}dq' }
    \left. +\hSig_{n+1,n}^{j,q} \rho_a \hSig_{n,n+1}^{j',q'} - \hSig_{n+1,n}^{j',q'} \rho_a \hSig_{n,n+1}^{j,q} \right)\nonumber\\
    &+&\sum_{j,j'} \sum_n \Gamma_n \int_{-\infty}^{\infty}dq \int_{-\infty}^{\infty}dq' 
    \left( \hSig_{n+1,n}^{j,q} \rho_a \hSig_{n,n+1}^{j',q'} + \hSig_{n+1,n}^{j',q'} \rho_a \hSig_{n,n+1}^{j,q} \right.\nonumber\\
    &&\hphantom{+\sum_{j,j'} \sum_n \int_{-\infty}^{\infty}dq \int_{-\infty}^{\infty}dq' }
    \left. -\hSig_{n,n+1}^{j,q}\hSig_{n+1,n}^{j',q'}\rho_a - \rho_a \hSig_{n,n+1}^{j',q'}\hSig_{n+1,n}^{j,q} \right).\nonumber\\
    \label{eqn:me_eff_2}
\end{eqnarray}

The master equation can be considerably simplified now because the integrals over $q,q'$ no longer involve $\chi$ and $\Gamma$. By interchanging the dummy variables $(j,q)\leftrightarrow(j',q')$ terms in the third line cancel. Also, the two terms on the second line can be cast in a Hamiltonian form. We then transform to an interaction picture with free evolution Hamiltonian 
\begin{equation}
     \hat{H}_f = \sum_j \sum_n \int_{-\infty}^{\infty}dq \; 4\hbar\omega_r\left(n^2+ \tdq^2\td{\sigma}_q^2 \right)\hat{\sigma}_{n,n}^{j,q},
\end{equation}
and denote the interaction picture density matrix by $\td{\rho}_a$, to arrive at the effective master equation 
\begin{eqnarray}
    \dot{\td{\rho}}_a &=& \frac{1}{i\hbar}\left[\hat{H}_\text{eff}, \td{\rho}_a \right]  
    +\sum_{j,j'}\sum_n\Gamma_n\int_{-\infty}^{\infty}dq \int_{-\infty}^{\infty}dq' \left( 2 \hat{\sigma}_{n+1,n}^{j,q} \td{\rho}_a \hat{\sigma}_{n,n+1}^{j',q'} \right. \nonumber\\
    &&\hphantom{+\sum_{j,j'}\sum_n \int_{-\infty}^{\infty}dq \int_{-\infty}^{\infty}dq'}
    \left.-\hat{\sigma}_{n,n+1}^{j',q'} \hat{\sigma}_{n+1,n}^{j,q} \td{\rho}_a 
    - \td{\rho}_a \hat{\sigma}_{n,n+1}^{j',q'} \hat{\sigma}_{n+1,n}^{j,q}  \right),\nonumber\\
    \label{eqn:me_eff_fin}
\end{eqnarray}

\noindent where the Hamiltonian $\hat{H}_\text{eff}$ is given by 
\begin{eqnarray}
    \hat{H}_\text{eff} &=& \sum_j \sum_n \int_{-\infty}^{\infty}dq \left(8n\hbar\omega_r\right)\left(\tdq\td{\sigma}_q\right)\hat{\sigma}_{n,n}^{j,q}\nonumber\\ 
    &+& \sum_{j,j'} \sum_n \hbar\chi_n \int_{-\infty}^{\infty} dq \int_{-\infty}^{\infty} dq' 
    \hat{\sigma}_{n,n+1}^{j,q}\hat{\sigma}_{n+1,n}^{j',q'}.\nonumber\\
    \label{eqn:h_eff}
\end{eqnarray}

\subsection{\label{sec:markov_validity}Validity of the Markov approximation}

Under the action of the Hamiltonian, Eq.~(\ref{eqn:h_eff_2_truncated}), squeezing proceeds at a rate $Q\sim N\chi_{\nDown}$ (assuming $\td{\sigma}_q \approx 0$) \cite{Ma2011PhyRep,swan2018PRL}. The Markov approximation used in Eq.~(\ref{eqn:bef_markov}) involves retaining only the leading term in the integration-by-parts expansion of the integrand. Neglecting the next-to-leading term amounts to approximating that 
\begin{equation}
    \left\lvert\frac{1}{\td{\rho}_a(t)}\frac{d\td{\rho}_a(t)/dt}{\kappa/2-i\delta_{\nDown}}\right\rvert \ll 1.
\end{equation}
Since the atomic dynamics proceeds at rate $\sim N\chi_{\nDown}$, the Markov approximation requires that $\abs{\kappa/2-i\delta_{\nDown}}\gg N\chi_{\nDown}$. 

\section{Implementing instantaneous state rotations}

\subsection{\label{sec:mcm_rot}Multi-center model}

In the multi-center model, we implement an instantaneous rotation in order to initialize the $c$-numbers in accordance with the initial state being in an equal superposition of the $\nDown,\nUp$ centers. We adopt a pragmatic approach to implement such a rotation: In the lab frame, we consider a fictitious Hamiltonian 
\begin{equation}
    \hat{H} = \frac{\Omega}{2} \sum_{j=1}^N \left(\hat{\sigma}_{\nDown,\nUp}^j e^{-i\theta} + \hat{\sigma}_{\nUp,\nDown}^j e^{i\theta}\right) 
\end{equation}
to act on the collection of atoms for a time $T=\pi/2\Omega$ so that the pulse area is $A=\pi/2$. Here $\theta$ specifies the orientation of the axis of rotation on the equatorial plane of the Bloch sphere. By ignoring the energy difference $\omega_m^j-\omega_n^j$ between any pair of states $n,m$, we are making the assumption that the pulse is `instantaneous'. While in practice any state preparation pulse requires a finite amount of time, here we assume such instantaneous pulses for simplicity and to avoid complications associated with pulse efficiencies and momentum widths \cite{Szigeti2012}.

\subsection{\label{sec:tcm_rot}Two-center model}
In the two-center model, instantaneous state rotations are used for state initialization and for probing the effect of echo pulses on the evolution of the squeezing parameter. To implement perfect, instantaneous rotations, we consider a Bloch sphere for each $\tdq$ value with the North and South poles represented by the states $\ket{\nUp,\tdq}$ and $\ket{\nDown,\tdq}$ respectively, since the rotation pulses do not couple states with different $\tdq$. The transformation of this pair of states under a rotation with axis $\mbf{\hat{n}}$ and pulse area $A$ ($\in [0,2\pi]$) is, 
\begin{equation}
    \begin{pmatrix}
    \ket{\nUp,\tdq}^\prime\\[5pt]
    \ket{\nDown,\tdq}^\prime
    \end{pmatrix}
    =
    U(\mbf{\hat{n}},A)
    \begin{pmatrix}
    \ket{\nUp,\tdq}\\[5pt]
    \ket{\nDown,\tdq}
    \end{pmatrix},
\end{equation}
where the matrix $U(\mbf{\hat{n}},A)$ is given by 
\begin{equation}
    U(\mbf{\hat{n}},A) = 
    \begin{pmatrix}
    \cos \frac{A}{2} - i n^z \sin \frac{A}{2}  &&  -i (n^x + i n^y) \sin\frac{A}{2}\\[5pt]
    -i (n^x - i n^y) \sin\frac{A}{2}   &&  \cos \frac{A}{2} + i n^z \sin \frac{A}{2}
    \end{pmatrix}.
    \label{eqn:u_mat}
\end{equation}
Since we track expectation values, we need to recast this transformation in terms of the means of one and two-atom operators. In what follows, we label $\nUp,\nDown$ using binary digits, i.e. $\nUp\equiv 0$ and $\nDown\equiv 1$. For one-atom operators, we define $\mbf{v}_1^{\tdq}$ with elements $v_1^{\tdq,j} = \ev{\hSig_{n_j,m_j}^{1,\tdq}}$, where $j=0,\ldots, 3$ and $n_j$ ($m_j$) is the second (first) digit from the right in the binary decomposition of $j$. The vector $\mbf{v}_1^{\tdq}$ transforms under the Bragg pulse to $\bar{\mbf{v}}_1^{\tdq} = M_1(\mbf{\hat{n}},A) \mbf{v}_1^{\tdq}$, where  
\begin{equation}
    M_1(\mbf{\hat{n}},A) = 
    \begin{pmatrix}
    \abs{U_{11}}^2  &&  U_{11}^*U_{21}  &&  U_{21}^*U_{11}  &&   \abs{U_{21}}^2\\[5pt]
    U_{11}^*U_{12}  &&  U_{11}^*U_{22}  &&   U_{21}^*U_{12}  &&   U_{21}^*U_{22}\\[5pt]
    U_{12}^*U_{11}  &&  U_{12}^*U_{21}  &&  U_{22}^*U_{11}  &&  U_{22}^*U_{21}\\[5pt]
    \abs{U_{12}}^2  &&  U_{12}^*U_{22}  &&  U_{22}^*U_{12}  &&  \abs{U_{22}}^2
    \end{pmatrix}.
\end{equation}
For two-atom operators, we similarly define $\mbf{v}_2^{\tdq,\tdq'}$ with elements $v_2^{\tdq,\tdq',j} = \ev{\hSig_{n_j,m_j}^{1,\tdq}\hSig_{r_j,s_j}^{1,\tdq}}$, where $j=0,\ldots, 15$ and $n_j,m_j,r_j,s_j$ are respectively the fourth, third, second and first digits from the right in the binary decomposition of $j$. This vector transforms as $\bar{\mbf{v}}_2^{\tdq,\tdq',j} = M_2(\mbf{\hat{n}},A) \mbf{v}_2^{\tdq,\tdq',j}$ where $M_2(\mbf{\hat{n}},A) = M_1(\mbf{\hat{n}},A) \otimes M_1(\mbf{\hat{n}},A)$ is a $16 \times 16$ matrix obtained as the Kronecker product of $M_1$ with itself.

\section{\label{sec:eom_cumulant}Evolution of expectation values of one and two-atom operators} 

We recall the dimensionless quantity $\tdq = q/\sigma_q$. For the numerical simulation, we consider $2L+1$ discrete $\tdq$ values to sample the Gaussian wavepacket within $r \sigma_q$ from center, where $r$ is a small natural number, typically $r=3$. As a result, we have \begin{equation}
    \tdq_j = \left(\frac{j}{L}-1\right) r, \; j=0,1,\ldots,2L.
\end{equation}
The one-atom expectation values are initialized as 
\begin{equation}
    \ev{\hSig_{\nDown,\nDown}^{1,\tdq}(0)} =  \frac{1}{\mathcal{N}}\frac{e^{-\tdq^2/2}}{\sqrt{2\pi}} \Delta \tdq,
\end{equation}
where $\Delta \tdq = r/L$ is the spacing between adjacent $\tdq$ values and the normalization constant 
\begin{equation}
\mathcal{N} = \sum_{j=0}^{2L} \frac{e^{-\tdq_j^2/2}}{\sqrt{2\pi}} \Delta \tdq 
\end{equation}
ensures that the norm of the initial density matrix is unity even with a finite number of samples. The two-atom expectation values are initialized as 
\begin{equation}
    \ev{\hSig_{\nDown,\nDown}^{1,\tdq} \hSig_{\nDown,\nDown}^{2,\tdq'} (0)} = \ev{\hSig_{\nDown,\nDown}^{1,\tdq}(0)} \ev{\hSig_{\nDown,\nDown}^{1,\tdq'}(0)}. 
\end{equation}

For the one-atom operators, the evolution of the expectation value is given by the following equation.

\begin{eqnarray}
    \frac{d}{dt} \ev{\hSig_{n_a,n_b}^{1,\tdq}} = &-&\left( \Gamma_{\nDown}\left(\delta_{n_a,\nDown} +\delta_{n_b,\nDown}\right) + i \chi_{\nDown}\left(\delta_{n_b,\nDown}-\delta_{n_a,\nDown}\right) + 8i\omega_r(n_b-n_a)\tdq \right) \ev{\hSig_{n_a,n_b}^{1,\tdq}} \nonumber\\
    &+& \delta_{n_a,\nUp}\delta_{n_b,\nUp} 2\Gamma_{\nDown}\ev{\hSig_{\nDown,\nDown}^{1,\tdq}} \nonumber\\
    &+& \delta_{n_b,\nUp}(N-1)\lambda_{\nDown}^* \sum_{j}\ev{\hSig_{n_a,\nDown}^{1,\tdq}\hSig_{\nDown,\nUp}^{2,\tdq_j}} \nonumber\\
    &-& \delta_{n_b,\nDown}(N-1)\lambda_{\nDown} \sum_{j}\ev{\hSig_{n_a,\nUp}^{1,\tdq}\hSig_{\nUp,\nDown}^{2,\tdq_j}} \nonumber\\
    &+& \delta_{n_a,\nUp}(N-1)\lambda_{\nDown} \sum_{j}\ev{\hSig_{\nDown,n_b}^{1,\tdq}\hSig_{\nUp,\nDown}^{2,\tdq_j}} \nonumber\\
    &-& \delta_{n_a,\nDown}(N-1)\lambda_{\nDown}^* \sum_{j}\ev{\hSig_{\nUp,n_b}^{1,\tdq}\hSig_{\nDown,\nUp}^{2,\tdq_j}},
    \label{eqn:one_atom}
\end{eqnarray}

where $\lambda_n = \Gamma_n + i\chi_n$ and the index $j$ runs from $0$ to $2L$.  

The expectation values of two-atom operators are governed by the following equation.
{ \allowdisplaybreaks

\begin{eqnarray}
    \frac{d}{dt} \ev{\hSig_{n_a,n_b}^{1,\tdq} \hSig_{n_c,n_d}^{2,\tdq'}} = &-&\left( \Gamma_{\nDown}\left(\delta_{n_a,\nDown}+\delta_{n_b,\nDown}+\delta_{n_c,\nDown}+\delta_{n_d,\nDown} \right) \right.\nonumber\\
    &&\left.
    + i\chi_{\nDown}\left(\delta_{n_b,\nDown}-\delta_{n_a,\nDown}+\delta_{n_d,\nDown}-\delta_{n_c,\nDown}\right)   \right. \nonumber\\
    &&\left. + 8i\omega_r\left(\left(n_b-n_a\right)\tdq + \left(n_d-n_c\right)\tdq' \right) \right) \ev{\hSig_{n_a,n_b}^{1,\tdq} \hSig_{n_c,n_d}^{2,\tdq'}} \nonumber\\
    &-& \delta_{n_b,\nDown}\delta_{n_d,\nUp} \lambda_{\nDown}\ev{\hSig_{n_a,\nUp}^{1,\tdq} \hSig_{n_c,\nDown}^{2,\tdq'}}
    - \delta_{n_a,\nUp}\delta_{n_c,\nDown}\lambda_{\nDown}^* \ev{\hSig_{\nDown,n_b}^{1,\tdq} \hSig_{\nUp,n_d}^{2,\tdq'}} \nonumber\\
    &-& \delta_{n_b,\nUp}\delta_{n_d,\nDown}\lambda_{\nDown}\ev{\hSig_{n_a,\nDown}^{1,\tdq} \hSig_{n_c,\nUp}^{2,\tdq'}}
    - \delta_{n_a,\nDown}\delta_{n_c,\nUp} \lambda_{\nDown}^* \ev{\hSig_{\nUp,n_b}^{1,\tdq} \hSig_{\nDown,n_d}^{2,\tdq'}} \nonumber\\
    &+& \delta_{n_a,\nUp}\delta_{n_b,\nUp} 2\Gamma_{\nDown} \ev{\hSig_{\nDown,\nDown}^{1,\tdq} \hSig_{n_c,n_d}^{2,\tdq'}}
    + \delta_{n_c,\nUp}\delta_{n_d,\nUp} 2\Gamma_{\nDown}\ev{\hSig_{n_a,n_b}^{1,\tdq} \hSig_{\nDown,\nDown}^{2,\tdq'}} \nonumber\\
    &+& \delta_{n_b,\nUp}\delta_{n_c,\nUp} 2\Gamma_{\nDown} \ev{\hSig_{n_a,\nDown}^{1,\tdq} \hSig_{\nDown,n_d}^{2,\tdq'}}
    + \delta_{n_a,\nUp}\delta_{n_d,\nUp} 2\Gamma_{\nDown}\ev{\hSig_{\nDown,n_b}^{1,\tdq} \hSig_{n_c,\nDown}^{2,\tdq'}} \nonumber\\
    &+& \delta_{n_a,\nUp} (N-2)\lambda_{\nDown} \sum_j \ev{\hSig_{\nDown,n_b}^{1,\tdq}\hSig_{n_c,n_d}^{2,\tdq'}\hSig_{\nUp,\nDown}^{3,\tdq_j}} \nonumber\\
    &-& \delta_{n_a,\nDown} (N-2)\lambda_{\nDown}^* \sum_j \ev{\hSig_{\nUp,n_b}^{1,\tdq}\hSig_{n_c,n_d}^{2,\tdq'}\hSig_{\nDown,\nUp}^{3,\tdq_j}} \nonumber\\ 
    &+& \delta_{n_c,\nUp} (N-2)\lambda_{\nDown} \sum_j \ev{\hSig_{n_a,n_b}^{1,\tdq}\hSig_{\nDown,n_d}^{2,\tdq'}\hSig_{\nUp,\nDown}^{3,\tdq_j}} \nonumber\\   
    &-& \delta_{n_c,\nDown} (N-2)\lambda_{\nDown}^* \sum_j \ev{\hSig_{n_a,n_b}^{1,\tdq}\hSig_{\nUp,n_d}^{2,\tdq'}\hSig_{\nDown,\nUp}^{3,\tdq_j}} \nonumber\\ 
    &+& \delta_{n_b,\nUp} (N-2)\lambda_{\nDown}^* \sum_j \ev{\hSig_{n_a,\nDown}^{1,\tdq}\hSig_{n_c,n_d}^{2,\tdq'}\hSig_{\nDown,\nUp}^{3,\tdq_j}} \nonumber\\    
    &-& \delta_{n_b,\nDown} (N-2)\lambda_{\nDown} \sum_j \ev{\hSig_{n_a,\nUp}^{1,\tdq}\hSig_{n_c,n_d}^{2,\tdq'}\hSig_{\nUp,\nDown}^{3,\tdq_j}} \nonumber\\
    &+& \delta_{n_d,\nUp} (N-2)\lambda_{\nDown}^* \sum_j \ev{\hSig_{n_a,n_b}^{1,\tdq}\hSig_{n_c,\nDown}^{2,\tdq'}\hSig_{\nDown,\nUp}^{3,\tdq_j}} \nonumber\\ 
    &-& \delta_{n_d,\nDown} (N-2)\lambda_{\nDown} \sum_j \ev{\hSig_{n_a,n_b}^{1,\tdq}\hSig_{n_c,\nUp}^{2,\tdq'}\hSig_{\nUp,\nDown}^{3,\tdq_j}} \nonumber\\    
    \label{eqn:two_atom}
\end{eqnarray}
}
To close the set of equations, we factorize the three-atom expectation values as 
\begin{eqnarray}
    \ev{\hSig_{n_a,n_b}^{1,\tdq}\hSig_{n_c,n_d}^{2,\tdq'}\hSig_{n_e,n_f}^{3,\tdq''}} \approx &&\ev{\hSig_{n_a,n_b}^{1,\tdq}\hSig_{n_c,n_d}^{2,\tdq'}}\ev{\hSig_{n_e,n_f}^{1,\tdq''}} + \ev{\hSig_{n_c,n_d}^{1,\tdq'}\hSig_{n_e,n_f}^{2,\tdq''}}\ev{\hSig_{n_a,n_b}^{1,\tdq}} \nonumber\\
    &+& \ev{\hSig_{n_e,n_f}^{1,\tdq''}\hSig_{n_a,n_b}^{2,\tdq}}\ev{\hSig_{n_c,n_d}^{1,\tdq'}} - 2 \ev{\hSig_{n_a,n_b}^{1,\tdq}}\ev{\hSig_{n_c,n_d}^{2,\tdq'}}\ev{\hSig_{n_e,n_f}^{1,\tdq''}}. \nonumber\\
\end{eqnarray}
To speed up computation, we identify ``partial sums'' which are recurring summations that appear in the evaluation of the right-hand-side of Eq.~(\ref{eqn:one_atom}) and Eq.~(\ref{eqn:two_atom}) for each $\tdq,\tdq'$, and evaluate these partial sums only once per time step (See Appendix A in Ref. \cite{shankar2019PRA}).

\section{\label{sec:param_cons}Laser power and squeezing rate}

Here, we explain how the constraint imposed by  Eq.~(\ref{eqn:beta_constraint}) translates to requirements on the laser power and limits on the squeezing rate.

\subsection{Laser power requirements}
Experimentally, the steady-state photon number $\abs{\beta}^2$ in mode 1 is set by the power in the drive laser as 
\begin{equation}
\abs{\beta}^2 = \frac{\kappa}{\kappa^2/4 + \Delta_{cl}^2} \left(\frac{P}{\hbar\omega_l}\right) \approx \frac{4R^2 P}{\hbar\omega_l\kappa},
\end{equation}
where the approximation assumes $\delta_{\nDown}\gg \omega_r$ so that $\Delta_{cl} \approx \delta_{\nDown}$, and that the interactions are in the dispersive regime i.e. $R^2\ll 1$. From Eq.~(\ref{eqn:beta_constraint}) and the discussion following it, the required laser power range is 
\begin{equation}
    \frac{2500\kappa}{R^2} \leq \frac{P}{\hbar\omega_l}\leq \frac{\Delta_c^2}{100C\gamma R^2}.
\end{equation}

\subsection{Squeezing rate}
From Eq.~(\ref{eqn:coup_strength_main}), the rate of squeezing $Q$ is proportional to $\abs{\beta}^2$, and consequently, the input power $P$, as 
\begin{equation}
    Q \approx \abs{\beta}^2 \frac{\gamma^2\kappa}{8\Delta_c^2}N C^2 R \approx \left(\frac{P}{\hbar\omega_l}\right) \frac{NC^2\gamma^2 R^3}{2\Delta_c^2} ,
    \label{eqn:q_power_relation}
\end{equation}
where we have assumed $R^2\ll 1$. Therefore, $Q$ is constrained to the range 
\begin{equation}
    \left[1250 \; R \left(\frac{\kappa C \gamma}{\Delta_c^2}\right)\right]N C \gamma \leq Q \leq \frac{R}{200}NC\gamma.
    \label{eqn:q_constraint}
\end{equation}

\section{\label{sec:rabi_osc}Rabi oscillation model for population leakage}

We consider the case when $\td{\sigma}_q\approx 0$. The two spin states correspond to $\ket{\nDown}=\ket{0\hbar k}$ and $\nUp=\ket{2\hbar k}$. We assume that mode 2 is dominantly sourced by the coherence between $\nDown$ and $\nUp$ and neglect the fluctuating terms to simplify the equation for $\zeta$ (Eq.~\ref{eqn:dtwa_mf}) to \begin{equation}
    \frac{d}{dt}\zeta = -\left(\frac{\kappa}{2}-i\Delta_{cl}\right)\zeta + i\frac{\gE}{2}\sum_j s_{\nUp,\nDown}^j.
    \label{eqn:zeta_simp}
\end{equation} 
We transform to the rotating frame $s_{\nUp,\nDown}^j = \td{s}_{\nUp,\nDown}^j e^{4i\omega_r t}$, $\zeta = \td{\zeta}e^{4i\omega_r t}$. At short times ($N\Gamma_{\nDown}t \ll 1$), assuming that the state is prepared along the $x$-axis of the Bloch sphere in the $\nDown,\nUp$ manifold, $\td{s}_{\nUp,\nDown}^j \approx 1/2$ for all $j$. Then, using the fact that $\abs{\kappa/2-i\delta_{\nDown}} \gg N\chi_{\nDown}$, we can adiabatically eliminate $\td{\zeta}$ as 
\begin{equation}
    \td{\zeta} \approx \frac{ig_\text{eff}/2}{\kappa/2-i\delta_{\nDown}}\sum_{j}\td{s}_{\nUp,\nDown}^j \approx -\frac{g_\text{eff}}{\delta_{\nDown}}\frac{N}{4}, 
    \label{eqn:zeta_td}
\end{equation}
where in the last approximation we have assumed that $R=\kappa/2\delta_{\nDown}\ll 1$. As an example of population leakage, we consider the $\nUp\leftrightarrow n_{+1}$ transition. By symmetry, the same arguments hold true for the $\nDown\leftrightarrow n_{-1}$ transition. Assuming $s_{n_{+1},n_{+1}}$ is negligible, $s_{\nUp,\nUp}\approx 1/2$ and zero populations and coherences associated with $n_{+2}$, the equation for the coherence $s_{n_{+1},\nUp}$ reads 
\begin{equation}
    \frac{d}{dt} s_{n_{+1},\nUp}^j = 12i\omega_r s_{n_{+1},\nUp} + i \frac{N}{4\delta_{\nDown}}\left(\frac{g_\text{eff}}{2}\right)^2 e^{4i\omega_r t},
\end{equation}
where we have used the expression for $\zeta$ from Eq.~(\ref{eqn:zeta_td}). From Eq.~(\ref{eqn:coup_strength_main}), the combination $g_\text{eff}^2/4\delta_{\nDown}$ can be immediately identified as $\chi_{\nDown}$ for $R\ll 1$. Solving for $s_{n+1,\nUp}^j$ gives 
\begin{equation}
    s_{n_{+1},\nUp}^j = -\frac{N\chi_{\nDown}}{32\omega_r}\left(e^{4i\omega_r t}-e^{12i\omega_r t} \right).
    \label{eqn:up_p1_coh}
\end{equation}
Further, still neglecting the $n_{+2}$ center, we can arrive at an equation for the dynamics of the population in $n_{+1}$ as 
\begin{equation}
    \frac{d}{dt}s_{n_{+1},n_{+1}}^j = i\frac{g_\text{eff}}{2}\left(\zeta^* s_{n_{+1},\nUp} - \zeta s_{\nUp,n_{+1}} \right),
\end{equation}
which can be solved using Eq.~(\ref{eqn:zeta_td}) and Eq.~(\ref{eqn:up_p1_coh}) to give 
\begin{equation}
    s_{n_{+1},n_{+1}}^j \approx \frac{1}{4}\left(\frac{N\chi_{\nDown}}{8\omega_r}\right)^2 \left(1 - \cos 8\omega_r t \right).
    \label{eqn:p1_pop_full}
\end{equation}
This expression explains the oscillations at frequency $\sim 8\omega_r$ that can be seen in the populations at the $n_{\pm 1}$ centers in Fig.~\ref{fig:diffBeta_sameR}(c-d), while the peak value scaled to the number of atoms gives the analytic expression for $P_{n_{\pm1}}^{\text{max}}$ (Eq.~(\ref{eqn:pop_pm1})) plotted in Fig.~\ref{fig:diffBeta_sameR}(e).

From Eq.~(\ref{eqn:up_p1_coh}), the maximum magnitude of the coherence $s_{n_{+1},\nUp}^j$ is $N\chi_{\nDown}/16\omega_r$. In estimating the intracavity field, we assumed that it is sourced only by the $s_{\nUp,\nDown}$ coherence. This approximation is valid as long as 
\begin{equation}
    \left\lvert\frac{s_{n_{+1},\nUp}}{s_{\nUp,\nDown}}\right\rvert \ll 1 \implies N\chi_{\nDown} \ll 8\omega_r.
    \label{eqn:strong_drive}
\end{equation}
The breakdown of the approximation, Eq.~(\ref{eqn:strong_drive}), signals the strong driving regime, i.e. it is the regime where the squeezing rate $N\chi_{\nDown}$ becomes comparable to the relative detuning  $8\omega_r$ between the $\nDown\leftrightarrow\nUp$ and $\nUp\leftrightarrow n_{+1}$, $n_{-1}\leftrightarrow\nDown$ transitions.

\section{\label{sec:fss}Relative importance of free-space scattering}

Here, we analyze the relative importance of single-atom free-space scattering and collective superradiant decay in increasing the variance $V_\text{min}$ that enters Eq.~(\ref{eqn:wineland}). Since the squeezing is driven by a term $\sim \hat{J}^z \hat{J}^z$, the axis corresponding to the minimum variance orients towards the $z$-axis over time \cite{Ma2011PhyRep}. As a result, we can estimate the degrading effect of various diffusive processes by estimating the corresponding increase in $\left(\Delta J^z\right)^2$. 

\noindent\emph{Free-space scattering:} We assume that once a photon is scattered into free-space, the atom recoils in a random direction and is lost from the atomic cloud. The rate of emission for a single atom is $\gamma \left(g^2\abs{\beta}^2/4\Delta_c^2\right)$, where the term in parenthesis is the effective population in $\ket{e}$ as a result of the drive laser. Starting with an equal superposition of $\ket{g,\nDown}$ and $\ket{g,\nUp}$,  each such photon could have been scattered equally likely from these two states, and so we have (assuming $\gamma t \ll 1$) 
\begin{equation}
    \dot{N}_{\nDown}/N = \dot{N}_{\nUp}/N = - \frac{\gamma}{2} \left(\frac{g^2\abs{\beta}^2}{4\Delta_c^2}\right).
\end{equation}
Scattering from the $\nDown$ ($\nUp$) state of any single atom increases (decreases) $J^z$ by $1/2$, therefore, the increase in variance in a time $t$ is 
\begin{eqnarray}
    \frac{\left(\Delta J^z\right)^2}{N/4} 
    &=& \frac{4}{N} N \left(\frac{\gamma t}{2} \left(\frac{g^2\abs{\beta}^2}{4\Delta_c^2}\right) \left( \left(-1/2\right)^2 + \left(1/2\right)^2  \right) \right.\nonumber\\
    &&\left.- \left(\frac{\gamma t}{2} \left(\frac{g^2\abs{\beta}^2}{4\Delta_c^2}\right) (-1/2 + 1/2) \right)^2 \right) \nonumber\\
    &=& \gamma t \left(\frac{g^2\abs{\beta}^2}{4\Delta_c^2}\right).
    \label{eqn:fss_var}
\end{eqnarray}

\noindent\emph{Superradiant decay:} The Lindblad term $\propto \Gamma_{\nDown}$ in Eq.~(\ref{eqn:me_compact}) contributes the following time evolution for $\ev{\hat{J}^z}$:
\begin{equation}
    \frac{d}{dt} \ev{\hat{J}^z} = 2\Gamma_{\nDown} \ev{\hat{J}^- \hat{J}^+} = 2\Gamma_{\nDown}\left(\ev{\mbf{\hat{J}}\cdot\mbf{\hat{J}}} -\ev{\hat{J}^z\hat{J}^z} -\ev{\hat{J}^z} \right),
\end{equation}
where we have used Eq.~(\ref{eqn:jmjp_breakdown}). For our initial state, we have $\ev{\mbf{\hat{J}}\cdot\mbf{\hat{J}}} = N/2(N/2+1)$, $\ev{\hat{J}^z\hat{J}^z} = N/4$ and $\ev{\hat{J}^z}=0$, so that,
\begin{equation}
    \dot{N}_{\nUp} = -\dot{N}_{\nDown} = \Gamma_{\nDown}N^2/2,
\end{equation}
where $N_{\nUp} \approx N/2 + \ev{\hat{J}^z}$ and $N_{\nDown} \approx N/2 - \ev{\hat{J}^z}$. The above rates are valid for times such that $N\Gamma_{\nDown}t \ll 1$. We can identify a per-atom rate of emission as $\Gamma_{\nDown}N/2$. Each such photon increases $J^z$ by $1$, therefore, the increase in variance in a time $t$ is 
\begin{equation}
    \frac{\left(\Delta J^z\right)^2}{N/4} = \frac{4}{N} N \left( \frac{N\Gamma_{\nDown}t}{2}(+1)^2 - \left(\frac{N\Gamma_{\nDown}t}{2}\right)^2 \right) \approx 2 N\Gamma_{\nDown}t.
    \label{eqn:sup_var}
\end{equation}

From Eq.~(\ref{eqn:fss_var}) and Eq.~(\ref{eqn:sup_var}), the contribution of free-space scattering can be neglected compared to that of superradiant decay when  
\begin{equation}
    \gamma t \left(\frac{g^2\abs{\beta}^2}{4\Delta_c^2}\right) \ll 2 N\Gamma_{\nDown}t \implies R^2 \gg \frac{1}{NC}.
\end{equation}
Here, $R=\kappa/2\delta_{\nDown}$ is assumed to be $\ll 1$. As a result, when $R^2$ becomes comparable to the inverse collective cooperativity, free-space scattering can no longer be neglected. In the simulations presented in this paper, $N=10^3$, $C=1,10$, giving $NC=10^3,10^4$. As a result, $R \gg 0.032, 0.01$ respectively for the two values of $C$.  The values of $R$ we consider are in the range $0.025-0.2$, and therefore some of our parameter regimes (e.g. $R=0.025, C=1$) do not satisfy the preceding requirement. A more precise estimate of the squeezing parameter for such regimes requires the inclusion of free-space scattering. Nevertheless, in an experiment, increasing the total number of atoms leads to a larger product $NC$ and stronger suppression of free-space scattering at fixed $R$.

\section*{References}
\providecommand{\newblock}{}

\end{document}